\documentclass[aps,prb,reprint,twocolumn,floatfix,nofootinbib,longbibliography]{revtex4-2} 
\usepackage[utf8]{inputenc}
\usepackage{graphicx} 
\pdfoutput=1
\usepackage{epsfig,amsmath,amssymb,color,comment,physics}
\usepackage[makeroom]{cancel}
\usepackage[caption=false]{subfig}
\usepackage{mathrsfs}
\usepackage[countmax]{subfloat}
\usepackage[normalem]{ulem}
\usepackage[english]{babel}
\usepackage[title]{appendix}
\usepackage{dsfont}
\usepackage{ragged2e}
\usepackage{float}
\usepackage{titlesec}
\usepackage{bbold}
\usepackage{tensor}
\usepackage[bookmarks=true,colorlinks,linkcolor=OrangeRed,urlcolor=NavyBlue,citecolor=RoyalBlue]{hyperref}
\usepackage[dvipsnames]{xcolor}
\usepackage{orcidlink}
\usepackage{accents}
\hypersetup{colorlinks, linkcolor={blue!80!blue}, citecolor={blue!50!blue}, urlcolor={blue!80!blue}, linktocpage=false}
\usepackage[normalem]{ulem}
\newcommand{\stkout}[1]{\ifmmode\text{\sout{\ensuremath{#1}}}\else\sout{#1}\fi}

\usepackage{alphabeta} 
\newcommand{\varpsi}{\textit{\textpsi}} 

\begin{document}

\title{Entanglement and apparent thermality in simulated black holes}

\author{Iason A. Sofos$^{*}$}
\affiliation{School of Physics and Astronomy, University of Leeds, Leeds LS2 9JT, UK}

\author{Andrew Hallam}
\affiliation{School of Physics and Astronomy, University of Leeds, Leeds LS2 9JT, UK}

\author{Jiannis K. Pachos}
\affiliation{School of Physics and Astronomy, University of Leeds, Leeds LS2 9JT, UK}

\begin{abstract}
We investigate the apparent thermality of Hawking radiation in the semi-classical limit of quantum black holes using the mean-field limit of a chiral spin-chain simulator, which models fermions propagating on a black hole space-time in the continuum. In this free-theory regime, no genuine thermalisation occurs. Nevertheless, we show that a bipartition across the event horizon yields a reduced density matrix whose mode occupations follow an apparent thermal Fermi-Dirac distribution. In contrast, partitions away from the horizon do not exhibit such thermal distributions, reflecting the absence of thermal behaviour. Our results demonstrate that Hawking radiation appears thermal only with respect to horizon bipartitions in free theories, with genuine thermal behaviour expected to emerge only in the presence of strong interactions deep in the black hole interior, a study of which we leave for future work.
\end{abstract}

\date{\today}
\maketitle

\begingroup
\renewcommand{\thefootnote}{\fnsymbol{footnote}}
\footnotetext[1]{\href{mailto:py19isleeds.ac.uk}{py19is@leeds.ac.uk}}
\endgroup

\section{Introduction}
The theoretical study of quantum black holes has revealed a deep relationship between general relativity, thermodynamics, and quantum field theory \cite{Bardeen_et_al_1973}. By viewing the entropy of a black hole as a measure of how much information the event horizon obscures from an external observer, Bekenstein \cite{Bekenstein_1973, Bekenstein_1974} argued that it was natural to associate a black hole with a physical entropy. The physical equivalence between the laws of black hole mechanics and thermodynamics was solidified when Hawking demonstrated that the vacuum state of the space-time containing a black hole is a thermal state with a well-defined temperature known as the Hawking temperature, $T_H$ \cite{Wald_1994, Hawking_1971, Hawking_1975}. This is the Hawking effect, and it is derived by analysing the semi-classical behaviour of quantum fields in a space-time containing a black hole. 

Despite theoretical advancements in the study of quantum black holes, there remains several fundamental open questions, including whether, via black hole formation and evaporation, information of the black hole's internal content is erased (the black hole information paradox) \cite{Hawking_1976b, Strominger_Vafa_1996}; why black hole entropy appears to be independent of the number of quantum fields \cite{Jacobson_1994, Susskind_Uglum_1994, Sakharov_1967}; what quantum degrees of freedom can be attributed to black hole entropy \cite{Wald_1975, Wald_2001}. In an effort to illuminate our understanding of these questions, there have been significant efforts, theoretically and, more recently, experimentally \cite{Philbin_et_al_2008, Weinfurtner_et_al_2011, Nguyen_et_al_2015, de_Nova_et_al_2019, Kolobov_et_al_2021, Hu_et_al_2019, Viermann_et_al_2022, Shi_et_al_2023, Cromb_et_al_2025, Rhyno_et_al_2026}, towards realising physical analogues of black holes. Such a system was first proposed by Unruh, who argued that sonic analogues of black holes would experience a process analogous to the Hawking effect \cite{Unruh_1981, Unruh_1995}. Since then, a myriad of systems have been proposed to simulate not only the physics of black holes but also cosmological inflation, the early universe, and aspects of the standard model ~\cite{Barcel_2005, Volovik_2001b, Volovik_2003, Volovik_2001, Huhtala_Volovik_2002}.

Here, we consider a chiral spin-chain model \cite{Horner_Thesis, Horner_hallam_Pachos_2023, Horner_et_al_2023, Daniel_Hallam_Horner_Pachos_2024} that, in its continuum limit, due to the emergence of an effective Lorentz invariance in its low-energy limit, describes a theory of non-interacting Dirac fermions in a curved space-time geometry. This Lorentz invariance is naturally broken at the length scales of the spin-chain's lattice spacing as a consequence of the model's non-linear dispersion relation. We calculate the density of states of the fermion zero-modes in various sub-regions of the space-time describing a (1+1)-dimensional black hole, which is finite as a consequence of the non-linear dispersion relation \cite{Volovik_2001, Huhtala_Volovik_2002, Volovik_2003, Corley_Jacobson_1996, Corley_Jacobson_1998, Corley_1998, Corley_Jacobson_1999, Jacobson_Parentani_2007}. By assuming that the fermions inside the black hole are well described by a thermal state at the Hawking temperature, the density of states can be used to compute the entropy of the fermion zero-modes across different regions of the black hole space-time. This includes the interior, where the entropy diverges logarithmically as the lattice spacing approaches zero~\cite{Holzhey_1994, Vidal_Latorre_Kitaev_2003, Calabrese_Cardy_2004}, as well as the exterior.

In parallel, we employ numerical simulations to investigate the entanglement entropy of the chiral spin-chain following a bipartition of the lattice into two subsystems. Generally, we expect that the entropy of the fermion zero-modes agrees with the entanglement entropy of the chiral spin-chain if the state describing the partitioned system is thermal, as, in this case, the von Neumann entropy is equivalent to the thermodynamic entropy of the subsystem's microstates \cite{Wald_2001, Mukohyama_1999}. When the system is partitioned at the event horizon, we find that the entanglement entropy is well described by the analytic fermion zero-mode entropy of the black hole’s interior, suggesting that the ground state of the partitioned system is thermal. However, when the system is partitioned further from the horizon, we find that this agreement breaks down, indicating that the subsystem's ground state appears thermal only when partitioned near the event horizon \cite{Chandran_2020, Belfiglio_2025}. We verify this by studying the subsystem’s mode occupation, which reveals a thermal Fermi-Dirac distribution only for horizon partitions. This breakdown of thermality for arbitrary partitions indicates that no genuine thermalisation occurs in the free-theory regime, implying that the simulated black hole’s information is not entirely erased \cite{Hayden_Preskill_2007, Xenos_Anastopoulos_Terzis_2026}.

To extract the Hawking temperature from these thermal distributions, it is necessary to transform the mean-field Hamiltonian into a form with only nearest-neighbour interactions and a linear coupling profile. In this form, the Hamiltonian's continuum limit approximates both the entanglement Hamiltonian globally and the Dirac Hamiltonian in a Rindler space-time near the event horizon, and the effective entanglement temperature associated with the partition coincides with the Hawking temperature. In the continuum, this transformation corresponds to a change of coordinates that preserves the curvature of the underlying space-time; thus, the physical value of the Hawking temperature remains invariant.

\section{Lattice simulator of Dirac fermions in black hole background}

In this section, we introduce the chiral spin-chain model \cite{Horner_Thesis, Horner_hallam_Pachos_2023, Horner_et_al_2023, Daniel_Hallam_Horner_Pachos_2024}, for which, due to the emergence of an effective Lorentz invariance at low energies, the continuum limit Hamiltonian is described by that of massless Dirac fermions propagating on a $(1+1)$-dimensional curved space-time geometry. Thus, by simulating the chiral spin-chain Hamiltonian, we can probe the entropy and thermality of fermions on a black hole space-time.

The chiral spin-chain model, which describes a one-dimensional chain of $N$ interacting spin-1/2 spins, is defined by the Hamiltonian~\cite{Horner_Thesis, Horner_hallam_Pachos_2023, Horner_et_al_2023, Daniel_Hallam_Horner_Pachos_2024}
\begin{equation}
    H=\sum^N_{n=1}\left[-\frac{u}{2}(\sigma^x_n\sigma^x_{n+1}+\sigma^y_n\sigma^y_{n+1}) +\frac{v}{4}\chi_n \right],
    \label{Chiral Spin-Chain Hamiltonian}
\end{equation}
where $u,v\in \mathbb{R}$, $\{\sigma^x_n,\sigma^y_n,\sigma^z_n\}$ are the Pauli matrices acting on the $n$th spin, $\chi_n=\boldsymbol{\sigma}_n\cdot(\boldsymbol{\sigma}_{n+1} \times \boldsymbol{\sigma}_{n+2})$ is a spin chirality operator \cite{DCruz_Pachos_2005, Tsomokos_et_al_2008}, and $\boldsymbol{\sigma}_n=(\sigma^x_n,\sigma^y_n,\sigma^z_n)$.

Obtaining the continuum limit of the chiral spin-chain begins with a Jordan-Wigner transformation, which maps the spins to a system of interacting fermions. For the analysis considered in the present work, we assume weak fluctuations, so a mean-field theory approximation can be applied. This transforms the chiral spin-chain Hamiltonian of Eq.~\eqref{Chiral Spin-Chain Hamiltonian} to the free fermion Hamiltonian \cite{Horner_Thesis, Horner_hallam_Pachos_2023, Horner_et_al_2023, Daniel_Hallam_Horner_Pachos_2024}
\begin{equation}
    H_{\text{MF}}=\sum^N_{n=1}\left(-uc^\dagger_nc_{n+1}-\frac{iv}{2}c^\dagger_nc_{n+2}\right) + \text{H.c.},
    \label{Equation: Mean-Field Hamiltonian}
\end{equation}
where $c^\dagger_n$ and $c_n$ are fermionic creation and annihilation operators, respectively, satisfying the anti-commutation relations $\{c_n, c^\dagger_m\}=\delta_{nm}$ and $\{c_n,c_m\}=\{c^\dagger_n,c^\dagger_m\}=0$. 

\begin{figure}
    \centering
    \includegraphics[width=0.85\linewidth]{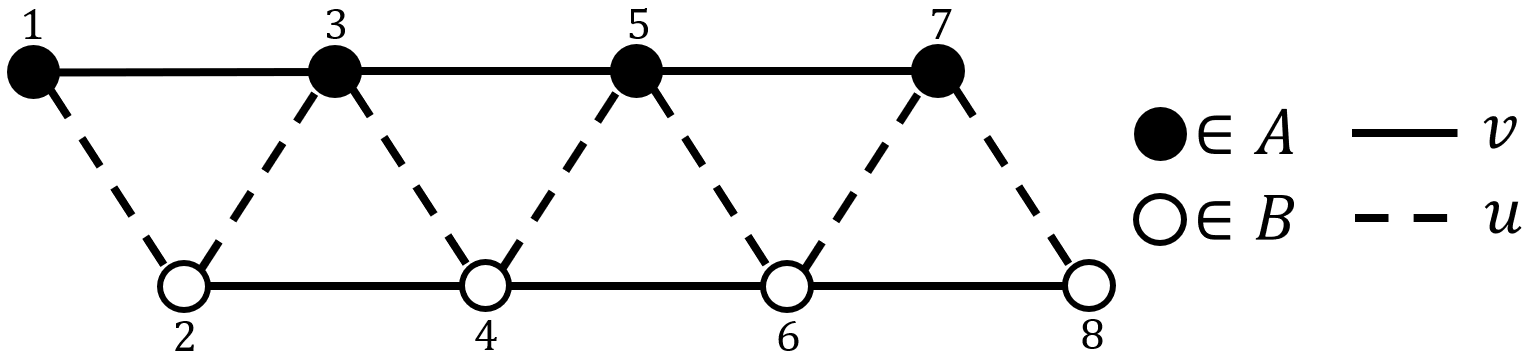}
    \caption{Schematic of the lattice corresponding to the chiral spin-chain model Hamiltonian of Eq.~\eqref{Chiral Spin-Chain Hamiltonian}, where the lattice sites alternate between two sub-lattices, $A$ and $B$, and a two-site unit cell has been introduced.}
    \label{Figure: Lattice Figure}
\end{figure}

\begin{figure*}
    \includegraphics[width=1\textwidth]{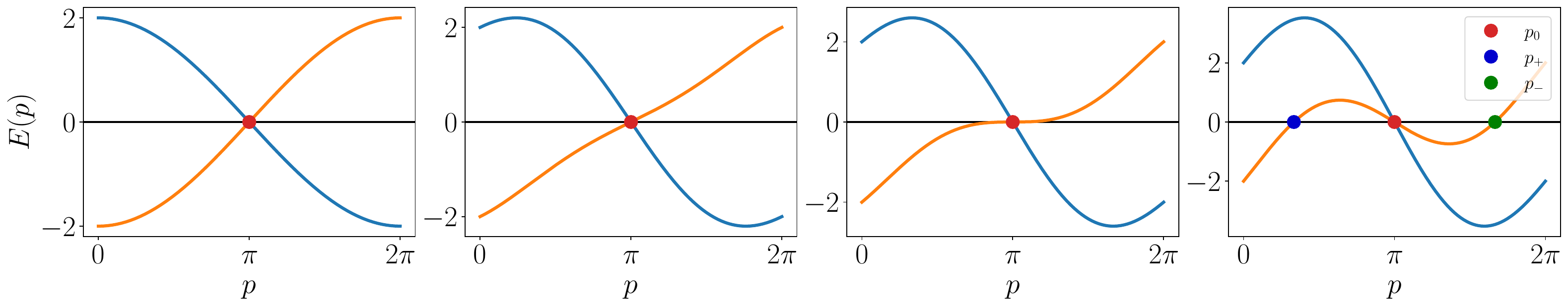}
    \centering
    \caption{Blue and orange curves show the positive (subluminal) and negative (superluminal) mean-field dispersion relations of Eq.~\eqref{Equation: MFT Hamiltonian Dispersion Relation}, respectively, for fixed $u=1$ and various coupling strengths $v$ (from left to right, $v=0, \ 0.5, \ 1, \ 2$). As the coupling $v$ increases, the Dirac cone about the Fermi point $p_0$ tilts. At the critical point $u^2=v^2$, which corresponds to the event horizon, the cone over-tilts, resulting in the emergence of two further Fermi points $p_\pm$.}
    \label{Figure: MFT Dispersion Relation Tilting}
\end{figure*}

By re-labelling the lattice sites such that they alternate between two sub-lattices, $A$ and $B$, and introducing a two-site unit cell, as illustrated in Fig.~\ref{Figure: Lattice Figure}, the mean-field Hamiltonian of Eq.~\eqref{Equation: Mean-Field Hamiltonian} can, following a Fourier transform, be expressed as \cite{Horner_Thesis, Horner_hallam_Pachos_2023, Horner_et_al_2023, Daniel_Hallam_Horner_Pachos_2024}
\begin{equation}
    H_{MF}=\sum_p\chi^\dagger_p h(p)\chi_p,
    \label{Equation: MFT Hamiltonian in Momentum Space}
\end{equation}
where $\chi_p=(a_p,b_p)^T$ is a two-component spinor and 
\begin{equation}
    h(p)=\begin{pmatrix} g(p)&f(p)\\ f^*(p)&g(p)\end{pmatrix}, 
    \label{Equation: Single Particle Hamiltonian}
\end{equation}
where $f(p)=-u(1+e^{-ia_cp})$ and $g(p)=v\sin(a_cp)$, is a single-particle Hamiltonian that can be diagonalised to obtain the following dispersion relation: \cite{Horner_Thesis, Horner_hallam_Pachos_2023, Horner_et_al_2023, Daniel_Hallam_Horner_Pachos_2024}
\begin{equation}
    E(p)=v\sin(a_cp)\pm 2u\cos\left(a_cp/2\right),
    \label{Equation: MFT Hamiltonian Dispersion Relation}
\end{equation}
where $a_c$ denotes the two-site unit cell spacing. This dispersion relation has Fermi points located at momenta $p_0=\frac{\pi}{a_c}$ and $p_\pm=\pm\frac{1}{a_c}\arccos(1-\frac{2u^2}{v^2})$, the latter of which exist only in the regime where $v^2>u^2$. In Fig.~\ref{Figure: MFT Dispersion Relation Tilting}, we see that the dispersion relation possesses Dirac cones that are tilted by the coupling $v$, resulting in the emergence of these Fermi points when these cones over-tilt \cite{Horner_Thesis, Horner_hallam_Pachos_2023, Horner_et_al_2023, Daniel_Hallam_Horner_Pachos_2024}.

The low-energy limit of the mean-field theory can be taken by Taylor expanding the single-particle Hamiltonian of Eq.~\eqref{Equation: Single Particle Hamiltonian} around the Fermi point $p_0$ to $O(p)$. Doing so gives 
\begin{equation}
    h(p+p_0)=-a_cv\mathbb{I}p+a_cu\sigma^yp\equiv e^x_a \alpha^a p,
\end{equation}
where the coefficients $e^x_0=-a_cv$ and $e^x_1=a_cu$ along with the matrices $\alpha^0=\mathbb{I}$ and $\alpha^1=\sigma^y$ have been defined, and $a\in\{0,1\}$. Following this, the continuum limit $a_c\rightarrow0$ and the thermodynamics limit $N\rightarrow \infty$ can be taken, whilst renormalising the couplings $a_cu\rightarrow u$ and $a_cv\rightarrow v$ and defining the spatial coordinate $x=na_c$ in the process. Thus, in this continuum and low-energy limit, the mean-field Hamiltonian of Eq.~\eqref{Equation: MFT Hamiltonian in Momentum Space} is given by \cite{Horner_Thesis, Horner_hallam_Pachos_2023, Horner_et_al_2023, Daniel_Hallam_Horner_Pachos_2024}
\begin{equation}
    H_{MF}=\int dp \chi^\dagger(p)e^x_a \alpha^a p\chi(p).
    \label{Equation: MFT Hamiltonian in Momentum Space (Continuum)}
\end{equation}

Finally, by performing a Fourier transform of the Hamiltonian of Eq.~\eqref{Equation: MFT Hamiltonian in Momentum Space (Continuum)} into real space, then a Legendre transformation, the following action $S$ can be obtained:
\begin{equation}
    S= i\int d^{1+1}x|e|\bar{\psi}(x)e^\mu_a\gamma^a\overset{\leftrightarrow}{\partial}_\mu\psi(x),
    \label{Equation: MFT Action}
\end{equation}
where $A\overset{\leftrightarrow}{\partial}_\mu B=\frac{1}{2}(A\partial_\mu B-(\partial_\mu A)B)$, the coefficients $e^t_0=1$ and $e^t_1=0$, the flat space-time gamma matrices $\gamma^a=\sigma^z\alpha^a$, and the conjugate spinor $\bar{\psi}(x)=\psi^\dagger(x)\gamma^0$ have been defined, $|e|=\det(e^a_\mu)$, and $\mu\in\{t,x\}$ \cite{Horner_Thesis, Horner_hallam_Pachos_2023, Horner_et_al_2023, Daniel_Hallam_Horner_Pachos_2024}.

The action of Eq.~\eqref{Equation: MFT Action} is that of a massless Dirac spinor $\psi$ on a (1+1)-dimensional Riemann-Cartan space-time, with the curved space-time spinor $\psi$ being related to the lattice spinor $\chi$ via the renormalisation $\chi=\sqrt{|e|}\psi$ and the zweibein being given by \cite{Horner_Thesis, Horner_hallam_Pachos_2023, Horner_et_al_2023, Daniel_Hallam_Horner_Pachos_2024}
\begin{equation}
    e^a_\mu = \begin{pmatrix} 1&v/u\\0&1/u\end{pmatrix}, \ \ 
    e^\mu_a=\begin{pmatrix}1&-v\\0&u\end{pmatrix},
    \label{Equation: Zweibein of GP Metric}
\end{equation}
which correspond to the following space-time metric line element: 
\begin{equation}
    ds^2=\left(1-\frac{v^2}{u^2}\right)dt^2-\frac{2v}{u^2}dtdx-\frac{1}{u^2}dx^2. 
    \label{Equation: Gullstrand-Painleve metric line element}
\end{equation}
It can be verified from the lattice spinor's anti-commutation relations and the zweibein of Eq.~\eqref{Equation: Zweibein of GP Metric} that the curved space-time spinor satisfies the necessary anti-commutation relations of Eq.~\eqref{Equation: Canonical Anti-Commutation Relations (Spinors)}.

If the couplings $v$ and $u$ are slowly varying position-dependent functions $v(x)$ and $u(x)$, then the line element of Eq.~\eqref{Equation: Gullstrand-Painleve metric line element} is that of the (1+1)-dimensional Gullstrand-Painlev\'e metric \cite{Gullstrand_1922, Painleve_1921}, which describes the space-time geometry of a (1+1)-dimensional black hole when the surface gravity $\kappa$ of the metric is positive, and a white hole when it is negative \cite{Volovik_2003, Volovik_2001} (see Appendix~\ref{Appendix: Surface Gravity of GP Metric}). The Gullstrand-Painlevé metric does not possess a coordinate singularity at the event horizon, which is located at $v(x_h)^2=u(x_h)^2$, and is therefore valid in both the interior and exterior space-time of the black/white hole, with the region bounded by $v(x)^2\geq u(x)^2$ corresponding to the interior \cite{Horner_hallam_Pachos_2023, Horner_Thesis}.  Though the Gullstrand-Painlevé metric is not \textit{static} (not invariant under the time-reversal transformation $t\mapsto-t$), it is \textit{stationary} (invariant under the time-translation transformation $t\mapsto t+t_0$). As the metric is stationary, it admits a time-like Killing vector $\xi^\mu=\delta^\mu_t$ (equivalently denoted $\partial_t$), which allows energy to be well-defined in the black hole's interior region and conserved along the metric’s geodesics \cite{Volovik_2001, Huhtala_Volovik_2002, Volovik_2003} (see Appendix~\hyperref[Appendix: Spinor fields on a (1+1)D Riemann-Cartan space-time]{A}).

It was demonstrated in Refs.~\cite{Horner_hallam_Pachos_2023, Daniel_et_al_2025} that a single-particle state initially localised in the region corresponding to the black hole's interior, which evolves unitarily under the mean-field Hamiltonian of Eq.~\eqref{Equation: Mean-Field Hamiltonian}, undergoes scattering across the event horizon that causes it to thermalise with a temperature given by $T_H$. The propagation of the single-particle state across the event horizon, which occurs due to mean-field theory's non-linear dispersion relation \cite{Volovik_2001, Huhtala_Volovik_2002, Volovik_2003, Corley_Jacobson_1996, Corley_1998, Corley_Jacobson_1999, Jacobson_Parentani_2007, Corley_Jacobson_1998}, is analogous to the quantum tunnelling interpretation of Hawking radiation \cite{Parikh_Wilczek_2000}.

\section{Entropy of simulated (1+1)D Dirac fermions in black hole background}
\label{Section: Entropy of (1+1)D Black Hole Fermions}
In this section, we derive analytic expressions for the entropy of the fermion zero-modes in the interior and exterior regions of a $(1+1)$-dimensional black hole. Two assumptions, that the state describing the fermions in the black hole's interior region is approximately thermal (that is, well described by a Gibbs state), and that the temperature of this state is the Hawking temperature, are key to this derivation. The derivation utilises a non-linear dispersion relation to find a finite density of states of the fermion zero-modes, which, in turn, can be used along with the assumptions of thermality to obtain their thermal energy and, hence, thermodynamic entropy. Under this assumption of thermality, we expect that the thermodynamic entropy of these fermion zero-modes is equivalent to the entanglement entropy associated with the bipartitioning of space-time by the event horizon \cite{Wald_2001, Mukohyama_1999}. In Sec.~\ref {Section: Entanglement entropy of chiral spin-chain simulator}, where numerical simulations of the chiral spin-chain's entanglement entropy are employed, this equivalence will be used to probe the thermality of the black hole.

In the mean-field description, the ground state corresponds to a filled Fermi sea with all negative-energy single-particle modes occupied. Fluctuations around this state are described by particles with positive energy and by antiparticles realised as holes created by removing fermions from the occupied negative energy states. Hence, in the low-temperature regime, the thermodynamic behaviour will be determined by excitations near the Fermi energy $E(p)=0$. The density of states $N(0)$, at zero energy, of the fermion zero-modes enclosed in a sub-region of length $s$ of the black hole's interior will be given by \cite{Volovik_2003, Volovik_2001}
\begin{equation}
    N(0)=\frac{N_F}{2\pi \hbar}\int^s_{a_c} dx\int_\mathbb{R}dp \ \delta (E(p)),
    \label{Equation: Density of States Integral}
\end{equation}
where $E(p)$ is the dispersion relation of the fermions and $N_F$ denotes the integer number of massless fermion fields.

As the effective Lorentz invariance emerges in the low-energy limit of the chiral spin-chain's continuum limit, we can identify the mean-field dispersion relation of Eq.~\eqref{Equation: MFT Hamiltonian Dispersion Relation} as an appropriate dispersion relation for the fermions in the black hole space-time. Moreover, as the continuum limit is defined microscopically on a lattice, the lattice spacing provides a natural ultraviolet cut-off that prevents the density of states of Eq.~\eqref{Equation: Density of States Integral} from being divergent. Operationally, this enters in two equivalent ways: First, the spatial integral should not be extended to separations smaller than $a_c$ from the event horizon $x_h$ and the coordinate origin $x=0$; that is, $a_c\leq s\leq x_h-a_c$ \cite{Bombelli_et_al_1986}. This prevents an infinite number of degrees of freedom, from points on either side of the horizon or the coordinate origin, from contributing to the density of states \cite{Bombelli_et_al_1986, Callan_1994, Holzhey_1994}. Second, as the mean-field dispersion relation is non-linear, it breaks the effective Lorentz invariance at distances comparable to the lattice spacing \cite{Volovik_2001, Huhtala_Volovik_2002, Volovik_2003, Corley_Jacobson_1996, Corley_1998, Corley_Jacobson_1999, Jacobson_Parentani_2007, Corley_Jacobson_1998}.  This allows for the density of states to be finite, as, by introducing a fundamental length scale, it prevents the localisation of an infinite number of degrees of freedom into an arbitrarily small volume \cite{Bombelli_et_al_1986, Callan_1994, Holzhey_1994}. 

In the regime corresponding to the black hole's interior, the mean-field dispersion relation of Eq.~\eqref{Equation: MFT Hamiltonian Dispersion Relation} has four zero-modes. When $u$ and $v$ are of the same sign, three of these zero-modes are that of the superluminal dispersion relation, located at momenta $p_0$ and $p_\pm$, and the fourth is that of the subluminal dispersion relation located at $p_0$. Thus, considering each of these zero-modes, and making use of the Dirac delta identity $\delta(f(x))=\sum_i \delta(x-x_i)/|f'(x_i)|$, the density of states of Eq.~\eqref{Equation: Density of States Integral} for the fermion zero-modes enclosed in the interior region of the black hole can be expressed as
\begin{widetext}
\begin{equation}
    N(0)=\frac{N_F}{2\pi \hbar}\int^{s}_{a_c}dx \left( \frac{1}{|1+v(x)|}+\frac{1}{|1-v(x)|}  +\Biggl|\frac{1}{\left(1-v(x)\right)}-\frac{1}{\left(1+v(x)\right)} \Biggr| \right),
    \label{Equation: Density of States for General Couplings}
\end{equation}
where the couplings have been renormalised and we have taken $u=1$, which can be solved for a given coupling $v(x)$ to obtain the relevant density of states. Here, we take the coupling $v(x)=\sqrt{x_h/x}$, which corresponds to that of a Schwarzschild black hole \cite{Volovik_2003, Volovik_2001}. For this coupling, the density of states of Eq.~\eqref{Equation: Density of States for General Couplings} can be solved to yield
\begin{equation}
    N(0)=\frac{2N_F x_h}{\pi \hbar}\left( \ln{\left| \frac{s^{\frac{1}{2}}+ x_h^{\frac{1}{2}}}{s^{\frac{1}{2}} -x_h^{\frac{1}{2}}} \right|} - \ln{\left|\frac{a_c^{\frac{1}{2}}+x_h^{\frac{1}{2}}}{a^{\frac{1}{2}}-x_h^{\frac{1}{2}}} \right|} -2\left(\frac{s}{x_h}\right)^\frac{1}{2} +2\left(\frac{a_c}{x_h}\right)^{\frac{1}{2}}\right).
    \label{Equation: Density of States Long and Exact Equation, in terms of s}
\end{equation}
\end{widetext}

Assuming that the reduced state associated with the fermions in the black hole's interior is well described by a thermal Gibbs state, the density of states can be used to find the thermal energy $\mathcal{E}(T)$ and, hence, the thermodynamic entropy $\mathcal{S}(T)$ of the fermions in the interior region of the black hole. The thermal energy of the fermions in the black hole's interior will be given by \cite{Volovik_2001, Huhtala_Volovik_2002, Volovik_2003}
\begin{equation}
    \mathcal{E}(T)=N(0)\int^\infty_0d{E}\frac{E}{e^{E/T}+1},
\end{equation}
which can be solved using the change of variables $u=-E/T$ and the dilogarithmic identity $\text{Li}_2(1)=\pi^2/6$ \cite{Morris_1979, Lewin_1958} to give \cite{Volovik_2001, Huhtala_Volovik_2002, Volovik_2003}
\begin{equation}
    \mathcal{E}(T)=\frac{\pi^2}{6}N(0)T^2,
    \label{Equation: Thermal Energy of Fermions}
\end{equation}
where $T$ is the temperature at which the fermions are in equilibrium. Using the first law of thermodynamics, $d\mathcal{E(T)}=Td\mathcal{S}(T)$, the entropy of the fermion zero-modes in the black hole's interior can be found to be
\begin{equation}
    \mathcal{S}(T)=\frac{\pi^2}{3}N(0)T.
    \label{Equation: Thermal Entropy of Fermions}\\
\end{equation}
Assuming that, in addition to being described by a thermal state, and as semi-classical intuition suggests, the temperature of this thermal state is equal to the Hawking temperature  \cite{Hawking_1971, Hawking_1975} 
\begin{equation}
    T_H=\frac{\hbar \kappa}{2\pi}= \frac{\hbar}{4\pi x_h},
    \label{Equation: Hawking Temperature of Chiral Spin-Chain Model}
\end{equation}
where $\kappa=-v'(x_h)$ is the surface gravity of the Gullstrand-Painlevé metric (see Appendix~\ref{Appendix: Surface Gravity of GP Metric}), the fermion zero-mode entropy of the black hole's interior region will be
\begin{equation}
    \label{Equation: Entropy of (1+1)D Black Hole Fermion Zero-Modes (Interior)}
    \begin{split}
    \mathcal{S}(T_H)&=\frac{N_F}{6}\Biggl( \ln{\Bigg| \frac{s^{\frac{1}{2}}+ x_h^{\frac{1}{2}}}{s^{\frac{1}{2}} -x_h^{\frac{1}{2}}} \Biggr|}- \ln{\Biggl|\frac{a_c^{\frac{1}{2}}+x_h^{\frac{1}{2}}}{a^{\frac{1}{2}} -x_h^{\frac{1}{2}}} \Biggr|} \\
     & \qquad \qquad \qquad -2 \left(\frac{s}{x_h}\right)^\frac{1}{2}+2\left(\frac{a_c}{x_h}\right)^{\frac{1}{2}}\Biggr).
    \end{split}
\end{equation}
Taking the length $s$ of the black hole's interior region to be $s=x_h-a_c$, such that the fermion zero-mode entropy of this region corresponds to that of the entire black hole, whilst working in the limit $0<a_c\ll x_h$, Eq.~\eqref{Equation: Entropy of (1+1)D Black Hole Fermion Zero-Modes (Interior)} for the fermion zero-mode entropy of the black hole's interior reduces to
\begin{equation}
    \mathcal{S}(T_H)=\frac{N_F}{6}\ln\left(\frac{4x_h}{a_c}\right)+\mathcal{S}_0,
    \label{Equation: Entropy of (1+1)D Black Hole Fermion Zero-Modes at Horizon}
\end{equation}
where $S_0=-N_F/3$, which is logarithmically divergent with the lattice constant $a_c$. It should be noted that, whilst the fermion zero-mode entropy Eq.~\eqref{Equation: Entropy of (1+1)D Black Hole Fermion Zero-Modes at Horizon} is akin to the entanglement entropy of a (1+1)-dimensional conformal field theory \cite{Holzhey_1994, Vidal_Latorre_Kitaev_2003, Calabrese_Cardy_2004}, caution should be taken in interpreting $N_F$ as the central charge, as we are considering a system with inhomogeneous couplings.

It is interesting to note that the black hole fermion zero-mode entropy given in ~\eqref{Equation: Entropy of (1+1)D Black Hole Fermion Zero-Modes at Horizon} exhibits a similar ultraviolet sensitivity familiar from entanglement entropy derivations of the Bekenstein-Hawking entropy. In gravitational theories, such divergences are typically absorbed via a renormalisation of the effective gravitational constant \cite{Jacobson_1994, Susskind_Uglum_1994, Jacobson_Parentani_2007}, along with the black hole entropy's dependence on the number of fermionic species \cite{Jacobson_1994, Susskind_Uglum_1994, Sakharov_1967}. In the present low-energy theory, the corresponding ultraviolet scale is the lattice spacing $a_c$.

In addition to studying the fermion zero-mode entropy of the black hole's interior, we can extend this analysis to derive the fermion zero-mode entropy of both the black hole's interior and exterior regions. The method of this derivation is analogous to that of the black hole's interior, and so, for brevity, we leave the details of this derivation to Appendix~\hyperref[Appendix: Entropy of Black Hole Exterior]{C}. It involves determining the density of states of Eq.~\eqref{Equation: Density of States for General Couplings} for both the interior region, which is bound by $[a_c,x_h-a_c]$, and a portion of the exterior space-time, which we take to be bound by $[x_h+a_c, s]$, where $s\geq x_h+a_c$ is the distance from the coordinate origin $x=0$. Here, as before, the lower limit of $s$ has been restricted to a distance $a_c$ from the event horizon to prevent degrees of freedom of points on either side of the event horizon with a separation of less than $a_c$ from contributing to the density of states. In contrast to before, only the two zero-modes corresponding to the Fermi point at $p_0$ will contribute to the density of states in the exterior region. Considering this, and once again assuming that the reduced state describing the fermions in the black hole's interior is suitably described by a thermal Gibbs state with a temperature $T_H$, the fermions' thermal energy and thermodynamic entropy can be defined, the latter of which can be solved to yield
\begin{equation}
    \mathcal{S}(T_H)=\frac{N_F}{6} \ln\left(\frac{4x_h}{a_c}\right)+\frac{\tilde{N}_F}{12}\left( \frac{s}{x_h} +\ln \left| \frac{s-x_h}{a_c}\right|\right)+\mathcal{S}'_0
    \label{Equation: Entropy of (1+1)D Black Hole Fermion Zero-Modes (Exterior)}
\end{equation}
in the limit $0<a_c\ll x_h$, where $\tilde{N}_F$ denotes the integer number of massless fermionic fields in the black hole space-time's exterior region and $\mathcal{S}'_0=\mathcal{S}_0-\tilde{N}_F/12$.

\section{Entanglement entropy of chiral spin-chain simulator}\label{Section: Entanglement entropy of chiral spin-chain simulator}
In the previous section, having assumed that the interior region is well described by a thermal state, we derived analytic expressions for the fermion zero-mode entropy for various regions of the black hole's space-time. We now numerically compute the entanglement entropy of the chiral spin-chain for arbitrary bipartitions. We do this for two distinct cases: first, when the partition is located at the event horizon; and second, when the partition is located at a lattice site in the region corresponding to the black hole's interior or exterior. We expect that the fermion zero-mode entropy agrees with the entanglement entropy if the ground state of the partitioned system is thermal, as, in this case, the von Neumann entropy equates to the thermodynamic entropy of the subsystem's microstates (fermion zero-modes) \cite{Wald_2001, Mukohyama_1999}. Thus, by comparing our analytic and numerical results, we can identify the regimes in which this equality holds and, by extension, the assumption of thermality is valid, and those in which it breaks down. We will see that as the system is free, the thermality condition is generally not valid, since, in the absence of interactions, the system cannot truly thermalise; we find that this equality holds only for event horizon partitions, due to the unique physics emerging in that region \cite{Hawking_1971, Hawking_1975}.

Consider a quantum system that is bipartitioned into two subsystems, $\mathcal{A}$ and $\mathcal{B}$, as illustrated in Fig.~\hyperref[Figure: Entropies for Interior, Horizon, Exterior, and n_s]{3a}, with a ground state $|\psi_{\mathcal{AB}}\rangle\in \mathcal{H_A}\otimes \mathcal{H_B}$ described by the density matrix $\rho_{\mathcal{AB}}=|\psi_{\mathcal{AB}}\rangle \langle \psi_{\mathcal{AB}}|$, where $\mathcal{H_A}$ and $\mathcal{H_B}$ denote the Hilbert spaces of $\mathcal{A}$ and $\mathcal{B}$, respectively. The bipartite entanglement entropy $S_\mathcal{A}$ of subsystem $\mathcal{A}$, which quantifies the correlations between $\mathcal{A}$ and $\mathcal{B}$, is defined as $S_\mathcal{A}=-\text{Tr}\left( \rho_\mathcal{A}\ln (\rho_\mathcal{A})\right)$, where $\rho_\mathcal{A}=\text{Tr}_\mathcal{B}(\rho_{\mathcal{AB}})=\sum_i\langle i_\mathcal{B}|\rho_{\mathcal{AB}}|i_\mathcal{B}\rangle$ denotes the reduced density matrix of $\mathcal{A}$ and $\{|i_{\mathcal{B}}\rangle \} \in \mathcal{H_B}$ the orthonormal basis of $\mathcal{H_B}$. 

\begin{figure}[t!]
    \centering
    \includegraphics[width=1\linewidth]{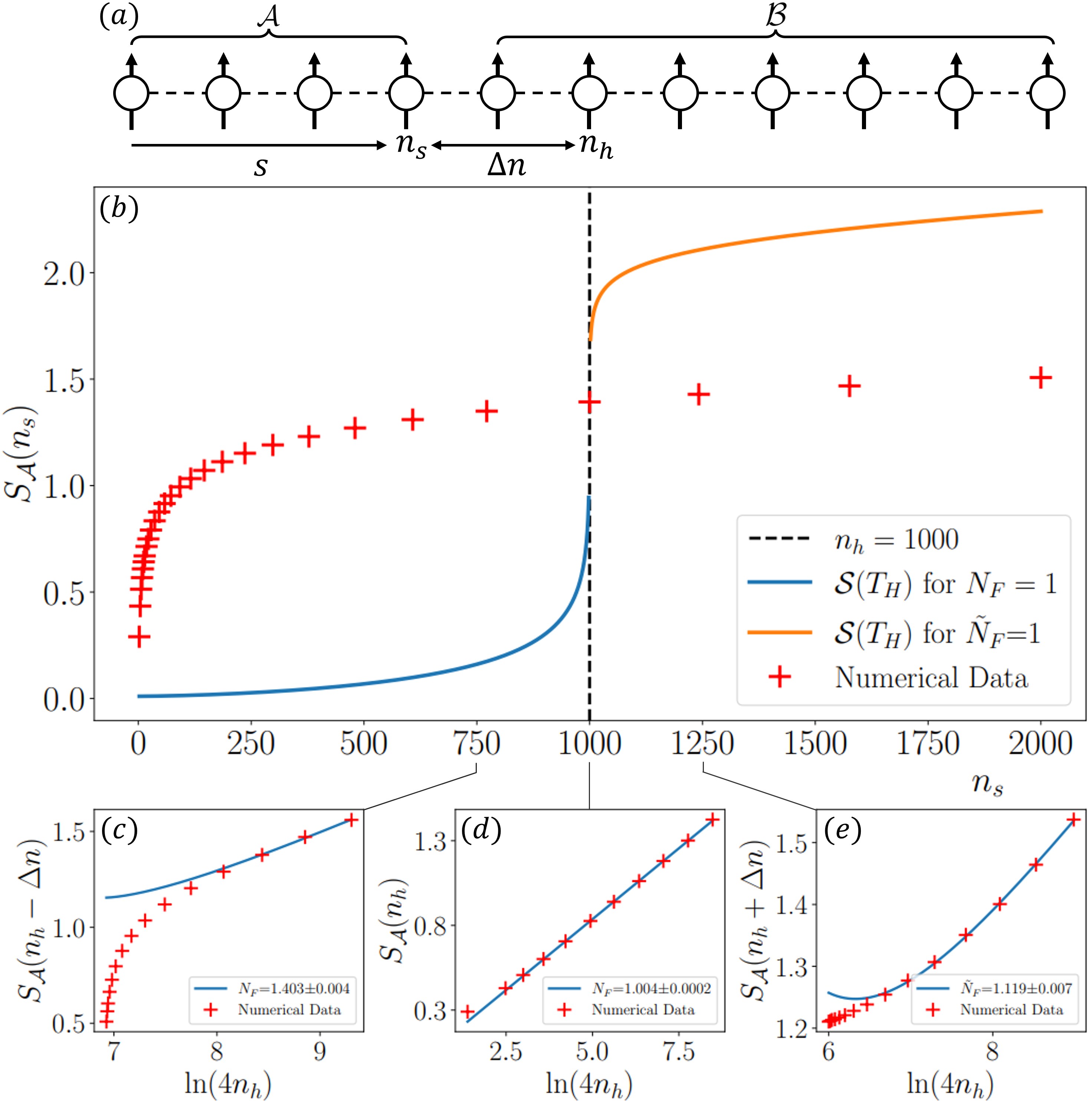}
    \caption{$(a)$ Schematic of spin-chain bipartition at lattice site $n_s$ located a distance $\Delta n$ from $n_h$. $(b)$ Red data points show entanglement entropy $S_{\mathcal{A}}{(n_s)}$ of $\mathcal{A}$ as a function of the partition site $n_s$ for a system with $n_h=1000$. Blue and orange curves show the fermion zero-mode entropies of Eqs.~\eqref{Equation: Entropy of (1+1)D Black Hole Fermion Zero-Modes (Interior)} and \eqref{Equation: Entropy of (1+1)D Black Hole Fermion Zero-Modes (Exterior)} for the black hole space-time's interior and exterior regions, respectively, with $N_F=\tilde{N}_F=1$. (c), (d), and (e) Red data points show the entanglement entropy $S_{\mathcal{A}}(n_s)$ of $\mathcal{A}$ as a function of the partition's location, where the system is partitioned at $n_s=n_h-\Delta n$ (interior), the event horizon $n_h$, and $n_s=n_h+\Delta n$ (exterior), respectively, with $\Delta n=250$. Blue curves show data interpolated using Eqs.\eqref{Equation: Entropy of (1+1)D Black Hole Fermion Zero-Modes (Interior)}, \eqref{Equation: Entropy of (1+1)D Black Hole Fermion Zero-Modes at Horizon}, and \eqref{Equation: Entropy of (1+1)D Black Hole Fermion Zero-Modes (Exterior)} for the fermion zero-modes' entropy for the interior, horizon, and exterior regions of the black hole space-time, respectively. To avoid partitioning the two-site unit cell of the continuum limit, only even $n_h$ were considered. For all figures, systems of $N=10000$ lattice sites with couplings $v(n)=\sqrt{n_h/n}$ and $u=1$ were taken; however, qualitative agreement between the entanglement and fermion zero-mode entropies persists until approximately 200 sites for horizon partitions.}
    \label{Figure: Entropies for Interior, Horizon, Exterior, and n_s}
\end{figure} 

We take the Hamiltonian of Eq.~\eqref{Equation: Mean-Field Hamiltonian} with couplings $v(n)=\sqrt{n_h/n}$ and $u=1$, which represent the Schwarzchild black hole profile, where we have introduced the discrete spatial coordinate $n=x/2a$. Here, $a=a_c/2=L/N$ denotes the spin-chain's lattice spacing, $L$ its physical length, and $n_h=x_h/a_c$ the lattice site where the horizon is located, defined as the site for which $v(n_h)=\pm1$. Throughout this work, we fix $L=1$ such that $x_h$ remains fixed in the thermodynamic $N\rightarrow\infty$ and continuum limit $a_c\rightarrow0$. As the Hamiltonian of Eq.~\eqref{Equation: Mean-Field Hamiltonian} is that of a free fermion system, the entanglement entropy can be efficiently computed via
\begin{equation}
    S_\mathcal{A}=-\sum_i\left(\lambda_i \ln(\lambda_i)+(1-\lambda_i)\ln(1-\lambda_i)\right),
\end{equation}
where $\lambda_i$ denotes the eigenvalues of the two-point correlation function $C_{ij}=\langle P_\mathcal{A}\phi_j|P_\mathcal{A}\phi_i\rangle$, $|P_\mathcal{A}\phi_i\rangle$ is the projection of the single-particle Hamiltonian's negative energy eigenstates onto $\mathcal{A}$, and $i,j\in \{1,...,N\}$ \cite{Peschel_2003, Klich_2005}. The entanglement entropy of this system with $N=10000$ lattice sites and event horizon located at $n_h=1000$ is given in Fig.~\hyperref[Figure: Entropies for Interior, Horizon, Exterior, and n_s]{3b}. 

We now turn to the analytic expressions for the entropy given in Eqs.~\eqref{Equation: Entropy of (1+1)D Black Hole Fermion Zero-Modes (Interior)}, \eqref{Equation: Entropy of (1+1)D Black Hole Fermion Zero-Modes at Horizon}, and \eqref{Equation: Entropy of (1+1)D Black Hole Fermion Zero-Modes (Exterior)}, which correspond to partitions in the interior, at the horizon, and in the exterior regions of the black hole's space-time, respectively, where we parametrise the partition's location by the integer $n_s=s/2a$ that is an absolute distance $\Delta n\geq 2$ from $n_h$. In these formulas, the integer parameters $N_F$ and $\tilde{N}_F$ remain to be determined. To fix these parameters, we numerically evaluate the entanglement entropy for a fixed $\Delta n$ as we vary the horizon's position $n_h$ and, hence, the partition $n_s$, as shown in Figs.~\hyperref[Figure: Entropies for Interior, Horizon, Exterior, and n_s]{3c}, ~\hyperref[Figure: Entropies for Interior, Horizon, Exterior, and n_s]{d}, and ~\hyperref[Figure: Entropies for Interior, Horizon, Exterior, and n_s]{e}. We find that, up to an overall additive constant, which is non-universal with respect to the chosen lattice parametrisation, only the expression corresponding to the horizon partition yields an integer value of $N_F \approx 1$, with this value converging as $N\rightarrow \infty$, whilst interior and exterior partitions suggest a larger value, both being in general non-integers. The convergence of $N_F\rightarrow 1$ in the thermodynamic limit, $N\rightarrow \infty$, for the horizon partition can be inferred from the entanglement entropy's derivative, which, as shown in Fig.~\ref{Figure: Interpolated Value of N_F for Varied delta_n, N=10000, N=1000, n_h=300},  plateaus around the value 1 for larger domains as the system size increases. In contrast to this, the deviation of $N_F$ from an integer value as the partition moves from the horizon is emphasised in Fig.~\ref{Figure: Interpolated Value of N_F for Varied delta_n, N=10000, N=1000, n_h=300}, where we plot its value from fitting the analytic formulas as $n_h$ varies for different $\Delta n$; we see that $N_F$ is not a constant, in contradiction to the assumptions used to determine the fermion zero-mode entropy.

To illustrate the entropy for general partitions $n_s \neq n_h$, we adopt $N_F = 1$ for the interior and $\tilde{N}_F = 1$ for the exterior, reflecting the value of the number of fermionic fields determined for a horizon partition. Moreover, we fix the non-universal constant of the exterior region's analytic zero-mode entropy to match the value obtained when partitioning at the horizon. As illustrated in Fig.~\hyperref[Figure: Entropies for Interior, Horizon, Exterior, and n_s]{3b}, we observe that the analytic zero-mode entropy formulas differ significantly from the numerically computed entanglement entropy in both the interior and exterior. Under the assumption of thermality, we expect the fermion zero-mode entropy to be equivalent to the entanglement entropy between the partitioned subsystems~\cite{Wald_2001, Mukohyama_1999}; thus, the increasing discrepancy between them as the partition deviates from the horizon signals the failure of thermality for general partitions. We also attribute the lack of consistent integer solutions for $N_F$ and $\tilde{N}_F$ to this breakdown of the assumed thermality that underpins the analytic derivations. In Sec.~\ref{Section: Mode Occupation and Black Hole Thermality}, this is verified by studying the subsystem's mode occupation, which reveals a Fermi-Dirac distribution only when the lattice is partitioned near the event horizon.

\begin{figure}[h!]
    \centering
    \includegraphics[width=0.9\linewidth]{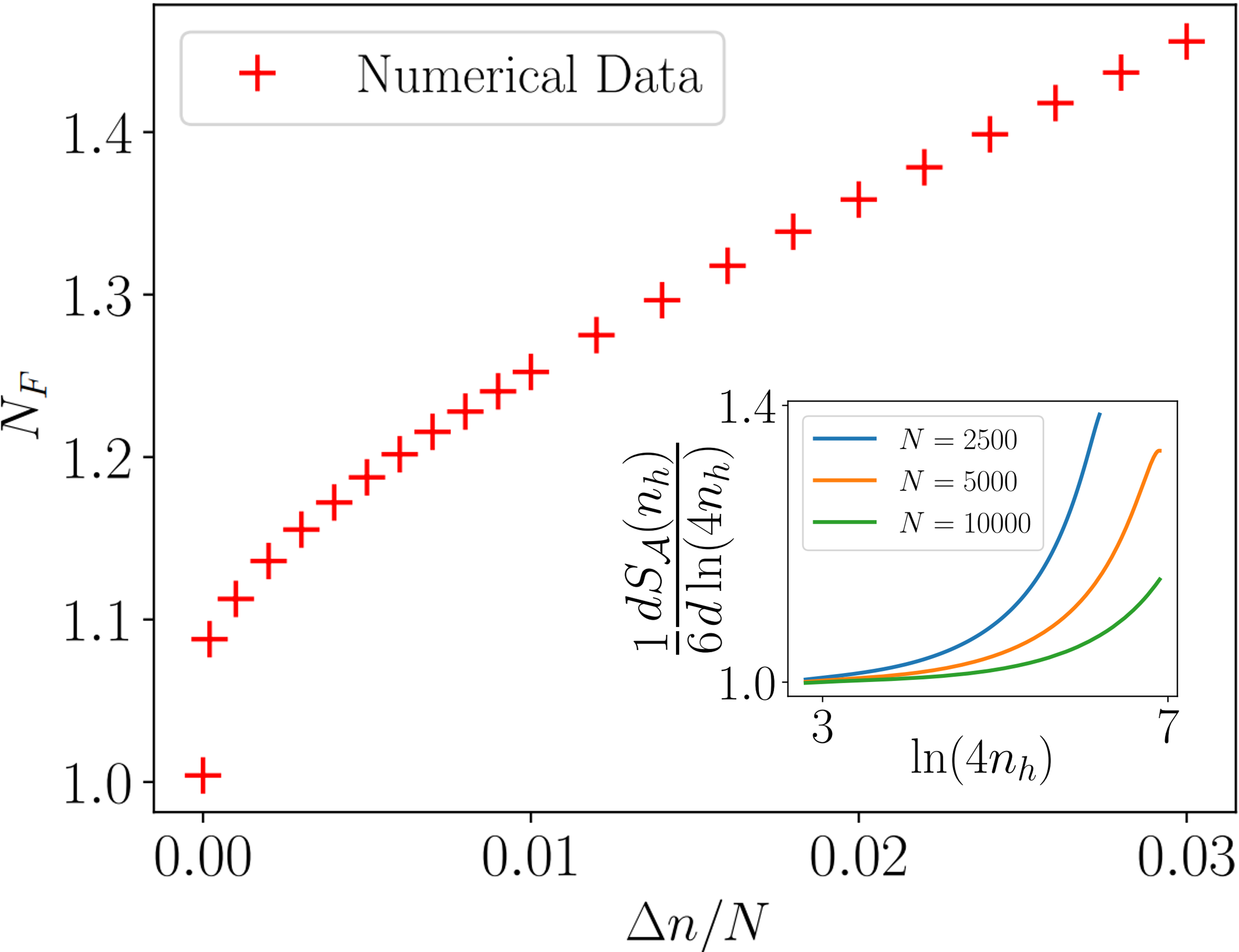}
    \caption{Data points show the interpolated value for the number of fermionic fields $N_F$ as a function of the distance $\Delta n$ of the partition $n_s=n_h-\Delta n$ from the event horizon $n_h$, for a system of $N=10000$ lattice sites with couplings $v(n)=\sqrt{n_h/n}$ and $u=1$. Each interpolated value of $N_F$ is determined by fitting Eq.~\eqref{Equation: Entropy of (1+1)D Black Hole Fermion Zero-Modes (Interior)} to numerical data for the entanglement entropy for a fixed $\Delta n$ as $n_h$ is increased. The deviation from $N_F\approx 1$ as $\Delta n$ increases indicates the breakdown of Eq.~\eqref{Equation: Entropy of (1+1)D Black Hole Fermion Zero-Modes (Interior)} for the fermion zero-mode entropy of the black hole's interior in describing the behaviour of the entanglement entropy as the partition deviates from the event horizon. Inset shows the derivative of the entanglement entropy $S_\mathcal{A}(n_h)$ with respect to $\ln(2n_h)$ following a horizon partition, $\Delta n=0$, for increasing system sizes, indicating the regime where $N_F\approx 1$. As system size increases, the domain for which the gradient plateaus around 1 increases, indicating that $N_F\rightarrow 1$ in the thermodynamic limit, $N\rightarrow\infty$.}
    \label{Figure: Interpolated Value of N_F for Varied delta_n, N=10000, N=1000, n_h=300}
\end{figure}

\section{Thermality of chiral spin-chain simulator}\label{Section: Mode Occupation and Black Hole Thermality}

\subsection{The breakdown of thermality}
To further investigate the breakdown of thermality for non-horizon partitions, we examine the mode occupation expectation value, $\langle 0_U| c_k^\dagger c_k|0_U\rangle$,  of the mean-field Hamiltonian that simulates the black hole space-time with respect to the ground state $|0_U\rangle$ of the asymptotically flat Hamiltonian (Eq.~\eqref{Equation: Mean-Field Hamiltonian} with couplings $v \rightarrow 0$ and $u = 1$). Here, $c_k$ are the eigenmodes that diagonalise the mean-field Hamiltonian of Eq.~\eqref{Equation: Mean-Field Hamiltonian} with the Gullstrand-Painlevé coupling profiles, $v(x)=\sqrt{x_h/x}$ and $u=1$, corresponding to a black hole space-time. The mode occupation can be expressed as \cite{Mertens_et_al_2022}
\begin{equation}
\langle 0_U| c^\dagger_k c_k|0_U\rangle=\sum_{q:E_U<0}|\langle E_{GP,k}|E_{U,q}\rangle|^2,
\label{Equation: Mode Occupation Expectation Value}
\end{equation}
where $|E_{GP,k}\rangle$ and $|E_{U,k}\rangle$ denote the $k$th eigenstate of the single-particle Hamiltonians, $h_{GP}$ and $h_U$, associated with $H_{MF}$ for the Gullstrand-Painlevé and asymptotically flat coupling profiles, respectively. Their corresponding energy eigenvalues are denoted $E_{GP,k}$ and $E_{U,k}$. Evaluating the mode occupation $\langle 0_U| c_k^\dagger c_k|0_U\rangle$ is the lattice equivalent to determining the Unruh vacuum expectation value perceived by an observer in a black hole space-time, which is the relevant quantity in the Hawking effect \cite{Mertens_et_al_2022, Wald_1994}, and should not be interpreted as a dynamical quench. The thermality that we probe here is thus that associated with the entanglement across the partition, rather than a post-quench locally measured temperature. 

\begin{figure}[h!]
    \centering
    \includegraphics[width=0.95\linewidth]{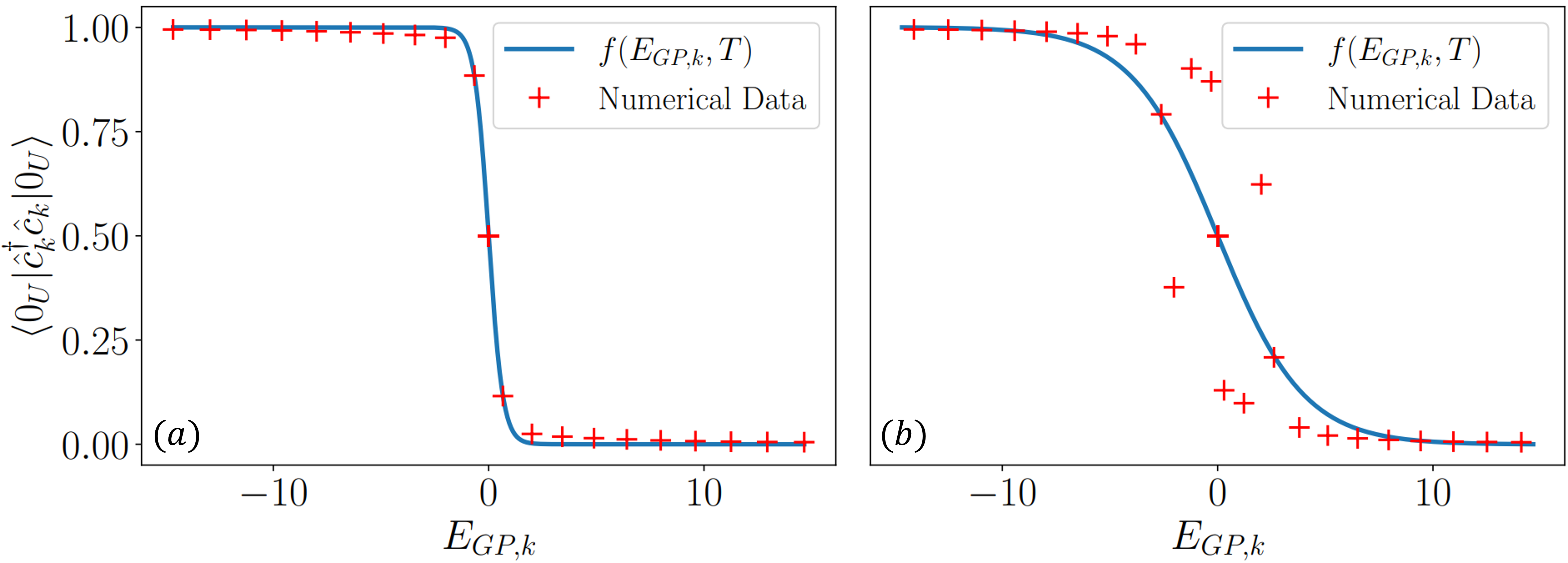}
    \caption{Red data points show the mode occupation expectation value $\langle 0_U| c_k^\dagger c_k|0_U\rangle$ of the mean-field Hamiltonian that simulates the black hole space-time (Eq.\eqref{Equation: Mean-Field Hamiltonian} with the Gullstrand-Painlev\'e couplings $v(x)=\sqrt{x_h/x}$ and $u=1$, defined only on $\mathcal{B}$) with respect to the ground state $|0_U\rangle$ of the asymptotically flat Hamiltonian (Eq.~\eqref{Equation: Mean-Field Hamiltonian} with couplings $v\rightarrow 0$ and $u=1$, defined over the entire lattice) for a horizon partition ($\Delta n=0$) and interior partition ($\Delta n=50$), respectively. The abscissa shows the energy eigenvalues $E_{GP,k}$ of the single-particle Hamiltonian $h_{GP}$ associated with $H_{MF}$ for the Gullstrand-Painlevé couplings; for finite $N$ the spectrum is bounded due to the lattice bandwidth. Blue curve shows data interpolated using the Fermi-Dirac distribution of Eq.~\eqref{Equation: Fermi-Dirac Distribution} with $T=0.327\pm0.004$ and $T=1.9\pm0.2$, and integrated interpolation errors $I_e=0.0003$ and $I_e=0.001$, respectively. System sizes of $N=2000$ lattice sites, with the horizon located at $n_h=700$, were considered.}
    \label{Figure: Fermi-Dirac for N=2000, n_h=699, delta_n=0}
\end{figure}

\begin{figure}[h!]
    \centering
    \includegraphics[width=0.95\linewidth]{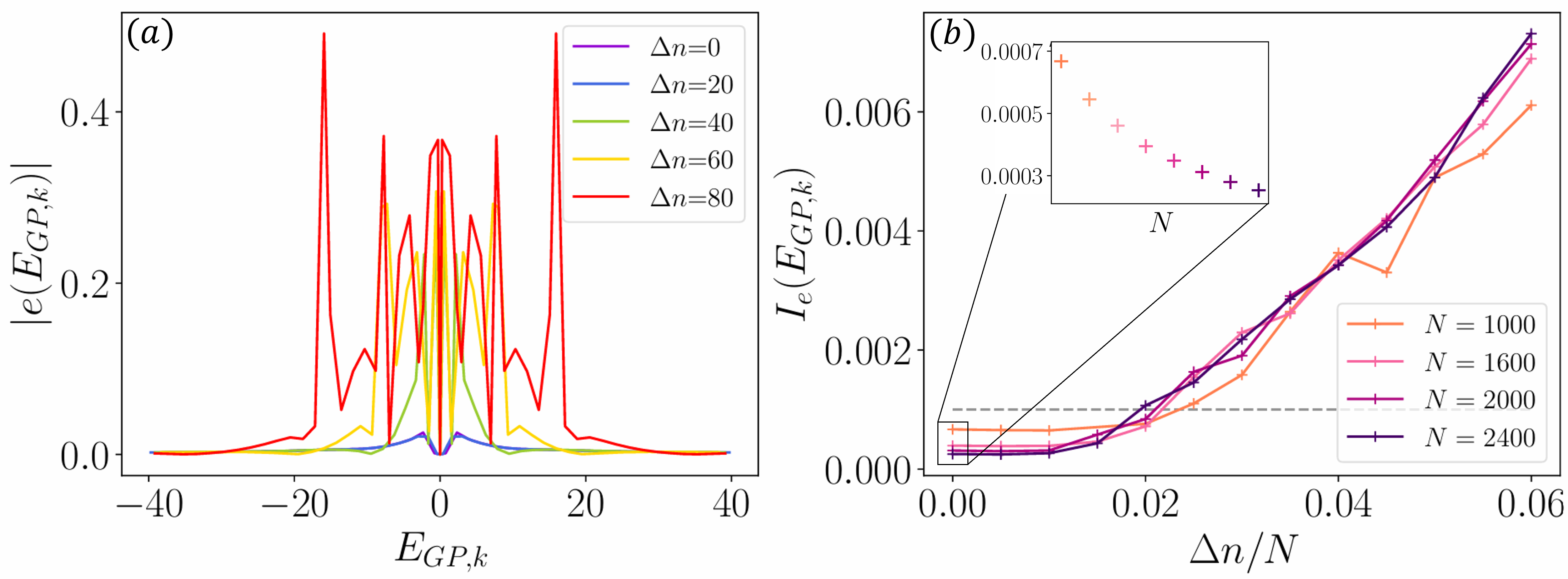}
    \caption{(a) Absolute value of interpolation errors $|e(E_{GP,k})|$ for the mode occupations $\langle 0_U| c_k^\dagger c_k|0_U\rangle$, interpolated using a Fermi-Dirac distribution, for systems of increasing $\Delta n$ with $N=2000$ lattice sites. As $\Delta n$ increases, and the system is partitioned further from $n_h$, increasing absolute interpolation errors indicate deviations from a Fermi-Dirac distribution in the mode occupation. (b) Integrated interpolation errors $I_e(E_{GP,k})$ for increasing $\Delta n$ and varied system sizes $N$, illustrating that the deviations from a Fermi-Dirac distribution as $\Delta n$ is increased are not minimised at larger system sizes. This contrasts with the horizon partition, shown in the inset, where the integrated interpolation errors decrease as system size increases. Dashed line indicates the maximum bound $I_e=0.001$ on distributions we consider thermal. For both figures, systems with $n_h=N/4$ were considered.}
    \label{Figure: Fermi-Dirac Interpolation Error for Varied delta_n}
\end{figure}

To perform the partial trace of subsystem $\mathcal{A}$, we define the Gullstrand-Painlevé Hamiltonian only in the region corresponding to subsystem $\mathcal{B}$ and take it to be zero elsewhere. This effectively traces out the degrees of freedom of $\mathcal{A}$. As illustrated in Fig.~\ref{Figure: Fermi-Dirac for N=2000, n_h=699, delta_n=0}, when the system is partitioned at the event horizon ($\Delta n = 0$), the mode occupation expectation value of Eq.~\eqref{Equation: Mode Occupation Expectation Value} follows a Fermi-Dirac distribution of the form 
\begin{equation}
    f(E_{GP,k},T)=\frac{1}{e^{E_{GP,k}/T}+1}, \label{Equation: Fermi-Dirac Distribution}
\end{equation}
where $T$ denotes the effective entanglement temperature associated with the partition, indicating that the reduced state is indeed thermal. However, as the partition is shifted away from the horizon ($\Delta n \neq 0$), deviations from a Fermi-Dirac distribution appear and grow with increasing $\Delta n$. This behaviour is quantified in Fig.~\ref{Figure: Fermi-Dirac Interpolation Error for Varied delta_n}, which shows the absolute and integrated interpolation errors, 
\begin{equation}
\begin{split}
    |e(E_{GP,k})|= |\langle&0_U|c_k^\dagger c_k|0_U\rangle - f(E_{GP,k}, T)|, \\
    I_e(E_{GP,k})&=\frac{1}{N}\sum_k|e(E_{GP,k}, T)|,
\end{split}
\end{equation}
respectively, that arise when fitting the mode occupation data to a Fermi-Dirac distribution. In contrast to the horizon partition, where these errors decrease with system size, these errors grow as $\Delta n$ increases, regardless of system size, confirming that the reduced state deviates progressively from thermal behaviour as the partition deviates from the horizon.  In particular, we see that the integrated interpolation error, whilst not exceeding $I_e=0.001$ for horizon partitions, readily increases for all system sizes after around $\Delta n/N\approx 0.02$. With this in mind, we posit that the integrated interpolation error can be used as a criterion for what we consider thermal. More concretely, we consider a reduced state to be thermal if the integrated interpolation error of the associated mode occupation does not exceeds $I_e=0.001$.

\subsection{Thermality and the Hawking temperature}
To extract an effective Hawking temperature from the simulated black hole system, the reduced density matrix of the free-fermionic ground state, which can be expressed as \cite{Peschel_Eisler_2009}
\vspace{0pt}
\begin{equation} 
    \rho_\mathcal{A}=\frac{e^{-H_e}}{Z},
    \label{Equation: Reduced Density Matrix} 
\end{equation}
where $H_e$ denotes the entanglement Hamiltonian and $Z=\mathrm{Tr}(e^{-H_e})$, must be equivalent to a thermal density matrix of finite temperature. For relativistic continuum theories, the Bisognano-Wichmann theorem \cite{Bisognano_Wichmann_1975, Bisognano_Wichmann_1976} relates the entanglement Hamiltonian of a half-space to the generator of Lorentz boosts via
\begin{equation} 
H_e = \int_{x\in\mathcal{A}} dx~\beta(x)\mathcal{H}(x), \qquad \beta(x)=2\pi x, 
\end{equation}
where $x$ measures the distance from the bipartition \cite{Dalmonte_2018} and $\beta$ is related to the effective entanglement temperature via $\beta=1/T$. Near the horizon, as the continuum geometry is locally described by the Rindler metric, tracing out the complementary region is analogous to restricting the vacuum to a Rindler wedge. Consequently, for a horizon bipartition, the reduced state is thermal with respect to the entanglement Hamiltonian, and the corresponding mode occupations display an apparent Fermi-Dirac distribution. Away from the horizon, as the bipartition no longer coincides with the causal horizon of the emergent geometry, the corresponding reduced density matrix is not expected to be thermal.

It was demonstrated in Ref.~\cite{Mertens_et_al_2022} that if, in addition to the near-horizon continuum geometry being locally Rindler, the lattice Hamiltonian approximates the entanglement Hamiltonian globally, then the effective entanglement temperature extracted from the Fermi-Dirac distribution is given by the Unruh temperature, which coincides with the Hawking temperature near the horizon. This occurs in free-fermionic systems with linear nearest-neighbour couplings, as due to the simultaneous alignment of the lattice Hamiltonian with the Rindler generator locally and the entanglement Hamiltonian globally, the mode occupations approximate those of the thermal Rindler vacuum.

We employ this method to probe the thermal properties of our chiral spin-chain simulator. However, the mean-field Hamiltonian of Eq.~\eqref{Equation: Mean-Field Hamiltonian} does not feature an overall linear coupling and, therefore, does not approximate the entanglement Hamiltonian globally, as is required to extract the Hawking temperature. To address this, the mean-field Hamiltonian must be modified such that the black hole geometry is encoded with only a linear nearest-neighbour coupling $u(x)$. In the continuum limit, such a modification corresponds to performing a coordinate transformation from Gullstrand-Painlevé to Schwarzschild coordinates, whilst preserving the space-time geometry (see Appendix \hyperref[Appendix: Conformal Invariance of Lattice Simulator]{D}); thus, by mapping the black hole geometry to a mean-field Hamiltonian that approximates the entanglement Hamiltonian, the mode occupations of the partitioned ground state yields a Fermi-Dirac distribution from which $T_H$ can be reliably extracted~\cite{Mertens_et_al_2022, Denger_Horner_Pachos_2023}. This procedure fails, however, when the chirality term $v(x)$ is non-zero, as in this case the mean-field Hamiltonian no longer approximates the corresponding entanglement Hamiltonian globally.

Consider the mean-field Hamiltonian of Eq.~\eqref{Equation: Mean-Field Hamiltonian} with a position-dependent nearest-neighbour coupling $u(x)$ and vanishing chirality term, $v(x) = 0$. In analogy with the chiral spin-chain, the low-energy continuum limit of this Hamiltonian corresponds to the action of a massless Dirac spinor on a $(1+1)$D curved space-time; the resulting geometry is described by the line element~\cite{Mertens_et_al_2022, Yang_2020}
\begin{equation}
    ds^2 = u(x){d\tilde{t}}^2 - \frac{dx^2}{u(x)},
    \label{Equation: Schwarzschild metric line element for XY-model}
\end{equation}
which matches the Schwarzschild metric in $(1+1)$ dimensions. In particular, for the choice $u(x) = 1-x_h/x$, this line element describes the exterior region of a Schwarzschild black hole with an event horizon located at $x_h$. This way, taking $v(x)$ to be zero and making $u(x)$ position dependent corresponds to a coordinate transformation from Gullstrand-Painlevé coordinates $(t, x)$ to Schwarzschild coordinates $(\tilde{t}, x)$. Whilst the Hamiltonian's structure changes under this transformation, scalar quantities, such as the space-time curvature, remain invariant.

Near the event horizon, the coupling $u(x)$ can be approximated by a linear profile
\begin{equation}
u(x) \approx 2\kappa(x - x_h),
\label{Equation: Linear Nearest-Neighbour Coupling u(x)}
\end{equation}
where $\kappa = \frac{1}{2}u'(x_h)$ is the surface gravity associated with the Schwarzschild metric. This expression for $\kappa$ can be obtained by mapping the Schwarzschild line element of Eq.~\eqref{Equation: Schwarzschild metric line element for XY-model} to a coordinate system that is regular at the horizon and applying the procedure detailed in Appendix~\ref{Appendix: Surface Gravity of GP Metric}. In this near-horizon regime, the Schwarzschild metric reduces to the Rindler metric, which describes the space-time of a uniformly accelerating observer in Minkowski space with a constant proper acceleration $\kappa$ (see Appendix~\ref{Appendix: Dirac Equation in Rindler Space-Time}).

According to the Fulling-Davies-Unruh effect~\cite{Fulling_1973, Davies_1975, Unruh_1975}, such an observer perceives their vacuum state to be a thermal state with temperature $T_H = \kappa / 2\pi$~\cite{Wald_1994} (see Appendices~\hyperref[Appendix: Hawking Effect in Chiral Black Hole Metric]{B}). It follows that the partitioned ground state of the mean-field Hamiltonian, with the linear coupling $u(x)$ of Eq.~\eqref{Equation: Linear Nearest-Neighbour Coupling u(x)} and $v=0$, will be thermal with an effective temperature equal to the Hawking temperature $T_H=\kappa/2\pi$, as it approximates both the entanglement Hamiltonian globally and the Hamiltonian of a Dirac spinor Hamiltonian in Rindler space-time locally near the horizon~\cite{Mertens_et_al_2022, Denger_Horner_Pachos_2023}. Here, tuning the gradient of $u(x)$ at the horizon corresponds to simulating the Hamiltonian of Rindler observers with different proper accelerations $\kappa$.

\begin{figure}[t!]
    \centering
    \includegraphics[width=0.95\linewidth]{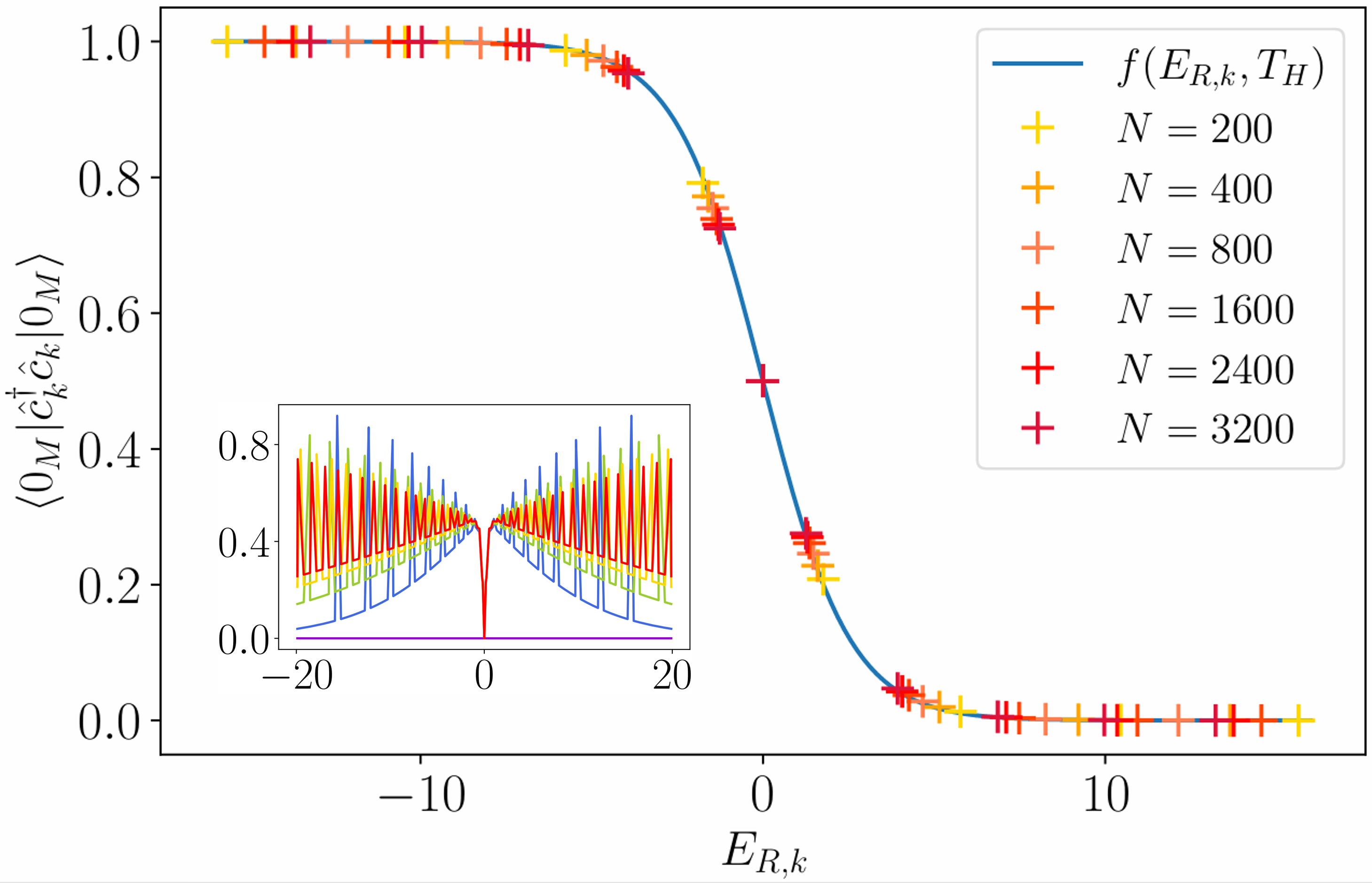}
    \caption{Data points show the mode occupation expectation values $\langle 0_M | c_k^\dagger c_k | 0_M \rangle$ of the mean-field Hamiltonian that simulates Rindler space-time (Eq.~\eqref{Equation: Mean-Field Hamiltonian} with couplings $u(x) = 2\alpha(x - x_h)$ and $v = 0$, defined only on $\mathcal{B}$), computed with respect to the Minkowski ground state $|0_M\rangle$ (Eq.~\eqref{Chiral Spin-Chain Hamiltonian} with $u = 1$ and $v = 0$, defined over the entire lattice). The abscissa shows the energy eigenvalues $E_{R,k}$ of the single-particle Hamiltonian $h_R$ corresponding to $H_{MF}$ with the Rindler couplings. The blue curves correspond to a Fermi-Dirac distribution with temperature $T_H = \kappa/2\pi$, where $\kappa = u'(x_h)/2 = \alpha$ is the surface gravity associated with the linear coupling profile. System sizes $N = 200$, $400$, $800$, $1600$, $2400$, and $3200$ were considered, with the horizon located at $n_h = N/2$ and coupling slope $\alpha = 2$. A minimal system size of $N\approx200$ was required to observe clean Fermi-Dirac distributions for horizon partitions. Inset depicts the absolute interpolation error $|e(E_{R,k})|$ between numerical and interpolated data of the mode occupation for $\Delta n = 0$, $20$, $40$, $60$, $80$ (colour scheme as in Fig.~\hyperref[Figure: Fermi-Dirac Interpolation Error for Varied delta_n]{6a}) with $N=2000$ and $\alpha=1/2$.}
    \label{Figure: XY Model Hawking Fermi Dirac Distribution}
\end{figure}

As in the Gullstrand-Painlevé case, the thermality of the Hamiltonian corresponding to a Dirac spinor in Rindler space-time can be confirmed by analysing the mode occupation of the partitioned system’s ground state. As Rindler space-time is globally flat, the relevant vacuum state for the Fulling-Davies-Unruh effect is that of the Minkowski vacuum. Thus, we compute the expectation values of the mode occupation operators with respect to the ground state $|0_M\rangle$ of the Minkowski Hamiltonian (Eq.~\eqref{Equation: Mean-Field Hamiltonian} with coupling profiles $v=0$ and $u=1$). As shown in Fig.~\ref{Figure: XY Model Hawking Fermi Dirac Distribution}, the resulting mode occupations for the Rindler Hamiltonian follow a Fermi-Dirac distribution with an effective entanglement temperature equal to $T_H$, thereby confirming the thermal nature of the reduced state and validating the extraction of the Hawking temperature via this approach. Once again, this thermality is only present for horizon bipartitions, as it is here where the emergent geometry coincides with the causal horizon.

\section{Discussion}
We studied the mean-field theory limit of a chiral spin-chain simulator that, in its continuum limit, due to the emergence of an effective Lorentz invariance at low energies, describes a theory of Dirac fermions on a black hole space-time. For this model, we obtained analytic expressions for the fermion zero-mode entropy in the interior and exterior regions of the black hole space-time. This entropy, derived under the assumption that the fermions are suitably described by a thermal Gibbs state with respect to any space-time bipartition, is finite as a result of the spin-chain's non-linear dispersion relation, which naturally breaks the effective Lorentz invariance at the length scales of the lattice spacing.

By employing numerical simulations to investigate the spin-chain's entanglement entropy and comparing it with our analytic black hole fermion zero-mode entropy, we identified regimes where the two entropies are equivalent and, by extension, where the assumption of thermality is valid and where it breaks down. Mainly, we demonstrated that, with respect to an arbitrary partition, the thermality condition is generally not valid. This is due to the absence of interactions, as the mean-field system cannot truly thermalise. Instead, the thermality condition holds only when the partition is located at the event horizon, due to the unique physics emerging in that region. We further verified this by studying the mode occupation of the partitioned system, which revealed a thermal Fermi-Dirac distribution only for partitions near the event horizon. 

When the chiral operator's coupling $v(x)$ is present, the mean-field Hamiltonian, which corresponds to that of a Dirac spinor on a Gullstrand-Painlevé space-time in its continuum limit, does not approximate an entanglement Hamiltonian. Hence, the temperature of this thermal distribution does not agree with the Hawking temperature. To extract the Hawking temperature, $v(x)$ must be set to zero, and the curvature needs to be encoded in a linear coupling $u(x)$ of the nearest-neighbour tunnellings. In the continuum limit, taking these couplings is equivalent to performing a coordinate transformation from Gullstrand-Painlevé to Schwarzschild space-time, which is approximated by Rindler space-time near the black hole's horizon. In this case, the mean-field Hamiltonian approximates both the entanglement Hamiltonian and the Hamiltonian of a Dirac spinor in Rindler space-time; thus, the Hawking temperature can be extracted from the thermal distribution of mode occupations. Once again, this thermal distribution breaks down as the partition deviates from the event horizon. 

The breakdown of thermality for arbitrary partitions indicates that no genuine thermalisation occurs in the free-theory regime, implying that the simulated black hole's information is not entirely erased. Beyond the mean-field limit, we expect strong interactions inside the simulated black hole to induce genuine thermalisation, resulting in a thermal distribution even for sufficiently small, but otherwise arbitrary partitions. Further investigation of thermalisation in this strong interaction regime is a potential avenue for future research, along with generalising the chiral spin-chain to higher-dimensional systems and exploring the interplay between Hamiltonian perturbations and coordinate transformation in the continuum limit.

\section{Conclusion}
We studied the mean-field theory limit of a chiral spin-chain simulator that, in its continuum limit, due to the emergence of an effective Lorentz invariance at low energies, describes a theory of Dirac fermions on a black hole space-time. For this model, we obtained analytic expressions for the fermion zero-mode entropy in the interior and exterior regions of the black hole space-time. This entropy, derived under the assumption that the fermions are suitably described by a thermal Gibbs state with respect to any space-time bipartition, is finite as a result of the spin-chain's non-linear dispersion relation, which naturally breaks the effective Lorentz invariance at the length scales of the lattice spacing.

By employing numerical simulations to investigate the spin-chain's entanglement entropy and comparing it with our analytic black hole fermion zero-mode entropy, we identified regimes where the two entropies are equivalent and, by extension, where the assumption of thermality is valid and where it breaks down. Mainly, we demonstrated that, with respect to an arbitrary partition, the thermality condition is generally not valid. This is due to the absence of interactions, as the mean-field system cannot truly thermalise. Instead, the thermality condition holds only when the partition is located at the event horizon, as here the emergent geometry is locally described by the Rindler metric and tracing out the complementary  region is analogous to restricting the system to a Rindler wedge. We further verified this by studying the mode occupation of the partitioned system, which revealed a thermal Fermi-Dirac distribution only for partitions near the event horizon. 

When the chiral operator's coupling $v(x)$ is present, the mean-field Hamiltonian, which corresponds to that of a Dirac spinor on a Gullstrand-Painlevé space-time in the continuum, does not approximate an entanglement Hamiltonian globally. Hence, the {effective entanglement temperature of this thermal distribution does not agree with the Hawking temperature. To extract the Hawking temperature, $v(x)$ must be set to zero, and the curvature needs to be encoded in a linear coupling $u(x)$ of the nearest-neighbour tunnellings. In the continuum limit, taking these couplings is equivalent to performing a coordinate transformation from Gullstrand-Painlevé to Schwarzschild space-time, which is approximated by Rindler space-time near the black hole's horizon. In this case, the mean-field Hamiltonian approximates both the entanglement Hamiltonian and the Hamiltonian of a Dirac spinor in Rindler space-time; thus, the Hawking temperature can be extracted from the thermal distribution of mode occupations. Once again, this thermal distribution breaks down as the partition deviates from the event horizon. 

The breakdown of thermality for arbitrary partitions indicates that no genuine thermalisation occurs in the free-theory regime, implying that the simulated black hole's information is not entirely erased. Beyond the mean-field limit, strong interactions might provide a mechanism for genuine thermalisation, resulting in thermal distributions for sufficiently small, but otherwise arbitrary partitions. Confirming this, however, remains an avenue for future research, along with generalising the chiral spin-chain to higher-dimensional systems and exploring the interplay between Hamiltonian perturbations and coordinate transformations in the continuum limit.

\section*{Acknowledgements}
We are grateful to Cristian Voinea, Matthew Horner, Aiden Daniel, and Tanmay Bhore for their insightful discussions and guidance. I.A.S. and J.K.P. acknowledge support from EPSRC with Grant Nos. EP/W524372/1 and UKRI1337:Anyons24, respectively. A.H. acknowledges support from Leverhulme Trust Research Leadership Award RL-2019-015. Statement of compliance with EPSRC policy framework on research data: This pub

\bibliography{main.bib}

\appendices
\renewcommand\thesection{\Alph{section}.}
\renewcommand\thesubsection{\arabic{subsection}}
\renewcommand{\theequation}{\thesection\arabic{equation}}
\setcounter{equation}{0}

\section{Spinor fields on a (1+1)D Riemann-Cartan space-time}\label{Appendix: Spinor fields on a (1+1)D Riemann-Cartan space-time}
This appendix is devoted to introducing the vielbein formalism of general relativity, which is the mathematical formalism required for dealing with spinors on a curved space-time. This formalism is required as the standard method for transitioning from special to general relativity, in which the Minkowski metric tensor $\eta_{\mu \nu}$ is replaced by the general covariant metric tensor $g_{\mu \nu}$ and partial derivatives $\partial_{\mu}$ (equivalently denoted $_{,\mu}$) by covariant derivatives $\nabla_\mu$ (equivalently denoted $_{;\mu}$), works only for objects that transform as tensors under Lorentz transformations, not for spinors, which are the mathematical objects that describe fermions \cite{Weinberg_1972}.

\subsection{Vielbein Formalism} \label{Subsection: Tetrad Formalism and Riemann-Cartan Geometry}
Consider an $n$-dimensional space-time $M$ with local coordinates $\{x^\mu \}$, where $x^\mu \in \{ t,x,y,z,\dots \}$ labels the coordinate axes, with coordinate basis vectors $\{e_\mu = \partial_\mu\}$ and dual basis vectors $\{ e^\mu = dx^\mu \}$ that span the cotangent space of $M$, which satisfy $e^\mu e_\nu = \delta ^{\mu}_{\nu}$ \cite{Parker_Toms_2009}. Greek indices $\mu,\nu,\dots \in \{t,x,y,z,\dots \}$ are used to denote components with respect to the coordinate basis $\{ e_\mu \}$. In the coordinate basis, the line element $ds^2$ of the space-time $M$ can be expressed as 
\begin{equation}
    ds^2 = g_{\mu \nu}(x)dx^\mu dx^\nu ,
    \label{Equation: Line element in coordinate basis}
\end{equation}
where $g_{\mu\nu}$ is the covariant metric tensor of space-time $M$ \cite{Parker_Toms_2009}. At any point $p$ on the space-time, the Principle of Equivalence can be used to construct a set of coordinates, spanned by the orthonormal basis vectors $\{ e_a \}$ with the corresponding dual basis vectors $\{ e^a\}$, which are locally flat at $p$ \cite{Weinberg_1972}. In such a coordinate system, the line element $ds^2$ is given by 
\begin{equation}
    ds^2 = \eta_{ab}e^a(x) e^b(x),
    \label{Equation: Line element in tetrad basis}
\end{equation}
where $\eta_{ab}$ is the Minkowski metric tensor \cite{Parker_Toms_2009}. As both $\{e^\mu  = dx^\mu\}$ and $\{e^a(x)\}$ span the cotangent space of $M$, it is possible to relate the two bases via
\begin{equation}
    e^a(x)=e^a_\mu (x) dx^\mu,
    \label{Equation: Relation between coordinate and vielbein bases}
\end{equation}
where the coefficients $e^a_\mu (x)$ are referred to as the vielbein (in two-, three- and four-dimensional space-times, these coefficients are often referred to as zweibein, dreibein and vierbein, respectively) \cite{Parker_Toms_2009}. Latin indices $a,b,\dots \in \{0,1,2,3,\dots \}$ are used to denote components with respect to the vielbein basis $\{e_a\}$.

Substitution of Eq.~\eqref{Equation: Relation between coordinate and vielbein bases} for the relation between the vielbein and coordinate bases into Eq.~\eqref{Equation: Line element in tetrad basis} for the line element as expressed in the vielbein basis, then comparing to Eq.~\eqref{Equation: Line element in coordinate basis} for the line element as expressed in the coordinate basis, yields \cite{Parker_Toms_2009}
\begin{equation}
    g_{\mu\nu}(x)=\eta_{ab}e^a_\mu(x) e^b_\nu(x),
    \label{Equation: Covariant Metric in Terms of Vielbein}
\end{equation}
which expresses the general covariant metric tensor in terms of the Minkowski metric and vielbein. The vielbein and their dual vectors, denoted $e^\mu_a(x)$, satisfy the relations \cite{Parker_Toms_2009}
\begin{equation}
    e^\mu_a(x)e^a_\nu(x)=\delta^\mu_\nu, \ \ \ \ e^a_\mu(x)e^\mu_b(x)=\delta^a_b,
\end{equation}
which, along with Eqs.~\eqref{Equation: Relation between coordinate and vielbein bases} and \eqref{Equation: Covariant Metric in Terms of Vielbein} for the definition of the vielbein and the general covariant metric tensor in terms of the vielbein, respectively, can be used to obtain 
\begin{equation}
    g^{\mu\nu}(x)=\eta^{ab}e^\mu_a(x)e^\nu_b(x).
    \label{Equation: Contravariant Metric in Terms of Vielbein}
\end{equation}

For a vector $\boldsymbol{X}$, the vielbein and their duals can be used to relate the vector's components in each basis via $X^\mu=e^\mu_a X^a$ and $X^a=e^a_\mu X^\mu$. The general covariant and contravariant metric tensors, $g_{\mu\nu}$ and $g^{\mu\nu}$, can be used to lower and raise coordinate indices, respectively. That is, $g_{\mu\nu} X^\nu = X_\mu$ and $g^{\mu\nu}X_\nu = X^\mu$. The covariant and contravariant Minkowski metric tensors, $\eta_{ab}$ and $\eta^{ab}$, can be used to lower and raise vielbein indices, respectively. That is, $\eta_{ab} X^b = X_a$ and $\eta^{ab}X_b = X^a$ \cite{Weinberg_1972}.

\subsection{Torsion, spin connection and Riemann-Cartan space-time}
To introduce a covariant derivative, which allows for the derivatives of tensors to be taken, a connection that describes how tensors should be transported around a space-time must be introduced. This enables tensors to be compared at infinitesimally separated points on the space-time, which allows the tensor's covariant derivative to be defined. The standard method for defining this connection is to impose that the lengths of, and angles between, a pair of vectors remain unchanged when being transported along a space-time \cite{Horner_Thesis}. This method of transporting vectors is known as parallel transport and requires that $d\boldsymbol{X}/d\lambda=0$, where $\lambda$ is some affine parameter.

The choice of connection defines the covariant derivative of a tensor. For example, the covariant derivative of a rank $(1,1)$ tensor $A^\mu_\nu$, expressed in terms of the coordinate basis, is given by
\begin{equation}
    \nabla_\alpha A^\mu_\nu = \partial_\alpha A^\mu_\nu +\Gamma^\mu_{\alpha\beta} A^\beta_\nu - \Gamma ^\beta _{\alpha\nu} A^\mu_\beta,
\end{equation}
where $\Gamma ^\alpha _{\beta \gamma}$ are the components of the connection. In standard general relativity, the connection that is chosen is the Levi-Civita connection $\Gamma ^\alpha_{\beta \gamma}=\Tilde{ \Gamma} ^\alpha_{\beta \gamma}$, which is completely determined by the metric and is given by \cite{Carroll_2019}
\begin{equation}
    \Tilde{\Gamma}^\alpha _{\beta \gamma}=\frac{1}{2}g^{\alpha \mu} \left( \partial_\gamma g_{\beta\mu} + \partial_\beta g_{\gamma\mu} - \partial_\mu g_{\beta\gamma} \right).
\end{equation}
The connection coefficients $\Tilde{\Gamma}^\alpha_{\beta\gamma}$ are often also referred to as the Christoffel symbols. The connection that is chosen for spinors is called the spin connection $\omega ^a_{\mu b}$ and, in the vielbein basis, is given by \cite{Farjami_et_al_2020}
\begin{equation}
    \omega^a_{\mu b}= e^a_\alpha\nabla_\mu e^\alpha_b=e^a_\alpha\left( \partial_\mu e^\alpha_b + \Gamma^\alpha_{\mu\beta}e^\beta_b \right),
\end{equation}
as it allows for the action that a covariant derivative has on a vielbein basis vector to be defined as $\nabla_\mu e_a=\omega^b_{\mu a}e_b$ \cite{Horner_Thesis}. Space-times with the spin connection are known as Riemann-Cartan space-times. The \textit{dreibein postulate} states that the covariant derivative of the dual dreibein $e^\nu _a$ is zero. That is, $\nabla_\mu e^\nu_a = 0$ \cite{Farjami_et_al_2020}. For this spin connection, the covariant derivative of a rank $(1,1)$ tensor $A^a_b$ in the vielbein basis is given by \cite{Farjami_et_al_2020}
\begin{equation}
    \nabla_\mu A^a_b = \partial_\mu A^a_b+\omega ^a_{\mu c}A^c_b - \omega ^c_{\mu b}A^a_c.
\end{equation}

\subsection{Spinor field on a Riemann-Cartan space-time}\label{Appendix: Spinor Field on a Riemann-Cartan Space-Time}
The action for a Dirac spinor $\psi$ of mass $m$ on a general (1+1)-dimensional Riemann-Cartan space-time $M$ is given by \cite{Nakahara_2003}
\begin{equation}
\begin{split}
    S_{RC}&=\int_Md^{1+1}x\mathcal{L}_{RC}\\
    &=\frac{i}{2}\int_Md^{1+1}x |e| \left( \bar{\psi}\gamma^\mu D_\mu \psi - \overline{D_\mu \psi} \gamma^\mu \psi +2im\bar{\psi}\psi \right),
    \label{Equation: Action for Dirac spinor of RC space-time (psi)}
\end{split}
\end{equation}
where $\mathcal{L}_{RC}$ is the Lagrangian density, $\gamma^\mu$ denotes the curved space-time gamma matrices that obey the anti-commutation relations $\{ \gamma^\mu, \gamma^\nu\} = 2g^{\mu\nu}$ and are related to the flat space-time gamma matrices $\gamma^a$ via $\gamma^\mu = e^\mu_a \gamma^a$, which themselves obey the anti-commutation relations $\{\gamma^a, \gamma^b\}=2\eta^{ab}$. The flat space-time gamma matrix $\gamma^0$ allows the Dirac adjoint $\bar{\psi}=\psi^\dagger \gamma^0$ to be defined. Here, $D_\mu$ denotes the covariant derivative of the spinor $\psi$ and $\overline{D_\mu\psi}$ denotes their adjoint, both of which are given by \cite{Farjami_et_al_2020}
\begin{align}
    D_\mu \psi = & \ \partial_\mu \psi +\omega _\mu \psi, \label{Equation: Spinor Covariant Derivative} \\
    \overline{D_\mu \psi} = \left(D_\mu \psi \right)^\dagger &\gamma^0= \partial_\mu\bar{\psi} - \bar{\psi}\omega_\mu, \label{Equation: Spinor Covariant Derivative Adjoint}
\end{align}
respectively, where $\omega_\mu$ is given by \cite{Farjami_et_al_2020}
\begin{equation}
    \omega_\mu = \frac{i}{2}\omega_{\mu a b}\Sigma^{ab}, \ \ \ \Sigma^{ab}=\frac{i}{4}[\gamma^a, \gamma^b],
    \label{Equation: Spin connection and Lorentz Algebra Generators}
\end{equation}
and $\omega_{\mu ab}=\eta_{ac}\omega^c_{\mu b}$. The notations $D_\mu$ and $\nabla_\mu$ are used to distinguish between the covariant derivative acting on spinors and tensors, respectively. 

From Eqs.~\eqref{Equation: Spinor Covariant Derivative} and \eqref{Equation: Spinor Covariant Derivative Adjoint} for the spinor covariant derivative and its adjoint, and the fact that $\{\omega_\mu,\gamma^\mu \}\propto \{\gamma^a,[\gamma^b,\gamma^c]\}=0$ in (1+1)-dimensions, Eq.~\eqref{Equation: Action for Dirac spinor of RC space-time (psi)} for the action of a massive Dirac spinor on a (1+1)-dimensional Riemann-Cartan space-time can be expressed as
\begin{equation}
    S_{RC}=i\int_M d^{1+1}x|e|\left( \bar{\psi}\gamma^\mu \overset{\leftrightarrow}{\partial}_\mu \psi +im\bar{\psi}\psi \right).
    \label{Equation: Action for Dirac spinor of (1+1) RC space-time (psi)}
\end{equation}

The canonical momentum $\pi(t,x)$ conjugate to $\psi(t,x)$ is defined as 
\begin{equation}
    \pi(t,x)=\frac{\partial\mathcal{L}}{\partial (\partial_t \psi)},
    \label{Equation: Canonical momentum definition}
\end{equation}
which, for a Dirac spinor on a general Riemann-Cartan space-time, is given by 
\begin{equation}
    \pi(t,x)=\frac{i}{2}|e|\psi^\dagger(t,x)\gamma^0\gamma^t.
    \label{Equation: Canonical Momentum for Dirac Spinor}
\end{equation}

To quantise the action for the Dirac spinor on a general Riemann-Cartan space-time given in Eq.~\eqref{Equation: Action for Dirac spinor of RC space-time (psi)}, $\psi(t,x)$ and $\pi(t,x)$ are promoted to Hermitian operators $\hat{\psi}(t,x)$ and $\hat{\pi}(t,x)$ and are required to satisfy the following equal-time canonical anti-commutation relations
\cite{Parker_Toms_2009}:
\begin{gather}
    \begin{split}
    \{\hat{\psi}_\alpha(t,x),\hat{\psi}_\beta(t,x')\}=\{\hat{\pi}&_\alpha(t,x),\hat{\pi}_\beta(t,x') \}=0, \\
    \{\hat{\psi}_\alpha(t,x), \hat{\pi}_\beta(t,x')\} &=i\delta_{\alpha\beta}\delta(x-x'),
    \end{split}
    \label{Equation: Canonical Anti-Commutation Relations (Canonical Momentum)}
\end{gather}
where $\alpha,\beta$ denote spinor components. Using these canonical anti-commutation relations and Eq.~\eqref{Equation: Canonical Momentum for Dirac Spinor} for the canonical momentum of a Dirac spinor, one obtains the following additional anti-commutation relation between the field operator $\hat{\psi}$ and its Hermitian conjugate $\hat{\psi}^\dagger$:
\begin{equation}
    \{\hat{\psi}_\alpha(t,x), \hat{\psi}_\beta^\dagger(t,x)\}=\frac{1}{|e|}(\gamma^0\gamma^t)^{-1}\delta_{\alpha\beta}\delta(x-x').
    \label{Equation: Canonical Anti-Commutation Relations (Spinors)}
\end{equation}

If the space-time $M$ is described by a stationary metric (that is, a metric that is invariant under the time-translation transformation $t\mapsto t+t_0$, where $t_0\in\mathbb{R}$), then it will admit a time-like Killing vector $\xi^\mu=\delta^\mu_t$ (often equivalently denoted $\partial_t$) satisfying the Killing equation of Eq~\eqref{Equation: Killing equation}. Consequently, the energy of the spinor will be a well-defined quantity that is conserved along the metric's geodesics \cite{Parker_Toms_2009, Huhtala_Volovik_2002}. This allows a Hamiltonian $H$ to be defined via the following Legendre transformation of the Lagrangian density: \cite{Farjami_et_al_2020}
\begin{equation}
    H=\int_Sdx\left( \frac{\partial \mathcal{L}}{\partial(\partial_t \psi)}\partial_t\psi-\mathcal{L} \right) ,
    \label{Equation: Legendre transform}
\end{equation}
Computing this Legendre transformation for the Lagrangian density of a spinor field in a (1+1)-dimensional Riemann-Cartan space-time yields the following Hamiltonian for a spinor on a Riemann-Cartan space-time: \cite{Parker_Toms_2009}
\begin{equation}
    \hat{H}=\int_Sdx\left(\hat{\pi}\partial_t\hat{\psi}\right).
    \label{Equation: Hamiltonian of Spinor on Riemann-Cartan Space-time}
\end{equation}

\section{Hawking effect in chiral black hole metric}\label{Appendix: Hawking Effect in Chiral Black Hole Metric}
\setcounter{equation}{0}
In this appendix, the Hawking effect is derived for a (1+1)-dimensional Schwarzschild black hole. The method of this derivation follows that of Refs.~\cite{Denger_Horner_Pachos_2023, Mertens_et_al_2022}, which utilises the equivalence between the Hawking effect and the Fulling-Davies-Unruh effect near the event horizon of a black hole. This equivalence is due to the fact that, near the event horizon, the Schwarzschild metric can be approximated by the Rindler metric, which describes a uniformly accelerating observer. This technique is used as it is more straightforward to analytically solve the Dirac equation and construct a quantum field theory in a space-time described by the Rindler metric than in a space-time described by the Schwarzschild or Gullstrand-Painlevé metrics. For brevity, the detailed mathematical construction of this quantum field theory in Rindler space-time is omitted, and the emphasis is instead on presenting the fundamental building blocks of this derivation. The final result of this section will be that an observer near the event horizon of the black hole will experience the Fulling-Davies-Unruh effect \cite{Fulling_1973, Davies_1975, Unruh_1975} with a temperature equal to the Hawking temperature of the black hole.

\subsection{Surface gravity of Gullstrand-Painlev\'e metric}\label{Appendix: Surface Gravity of GP Metric}
Consider the (1+1)-dimensional Gullstrand-Painlevé metric line element given in Eq.~\eqref{Equation: Gullstrand-Painleve metric line element}, which is the line element of the (1+1)-dimensional Schwarzschild metric expressed in Gullstrand-Painlevé coordinates $x^\mu=(t,x)$. Here, the couplings $u$ and $v$ are independent of $t$, but not necessarily of $x$; thus, the metric is stationary, though not static \cite{Volovik_2003}. As the Gullstrand-Painlev\'e metric is stationary, it admits a Killing vector $\xi^\mu=\delta ^\mu_t=(1,0)$ \cite{Parker_Toms_2009}, which is time-like in the region $u^2>v^2$ and satisfies the Killing equation
\begin{equation}
    \xi_{\mu;\nu}+\xi_{\nu;\mu}=0.
    \label{Equation: Killing equation}
\end{equation}
The vector $\xi_\mu=g_{\mu\nu}\xi^\nu=(1-v^2/u^2,-v/u^2)$ is also a Killing vector that satisfies the Killing equation. As the Gullstrand-Painlevé metric is regular at the event horizon $x_h$, these Killing vectors can be used to find the surface gravity $\kappa$ of the black hole space-time, as defined by a distant observer, by evaluating 
\begin{equation}
    \xi^\mu \xi_{\nu;\mu}=\xi^\mu \left(\xi_{\nu,\mu}- \xi_\gamma \tilde{\Gamma}^\gamma_{\nu\mu}\right)= \kappa \xi_\nu,
    \label{Equation: Surface gravity definition}
\end{equation}
at the black hole's event horizon \cite{Wald_1994, Wald_1984a}. 

Expanding the Einstein summations over $\mu$ and $\gamma$, and making use of the fact that $\xi_{\nu,t}=\forall\nu$, $\xi^t=1$, and $\xi^x=0$,  Eq.~\eqref{Equation: Surface gravity definition} for the surface gravity can be expressed as
\begin{equation}
    -\xi_x \tilde{\Gamma}^x_{\nu t} - \xi_t \tilde{\Gamma}^t_{\nu t}= \kappa \xi_\nu,
\end{equation}
which, using the only relevant and non-zero Christoffel coefficients $\tilde{\Gamma}^t_{xt}=\frac{1}{2} g^{tt}g_{tt,x}$ and\\
\noindent $\tilde{\Gamma}^x_{xt}=\frac{1}{2} g^{xt}g_{tt,x}$, can be solved for $\nu=x$ to yield
\begin{equation}
    -\frac{1}{2}\xi_tg^{tt}g_{tt,x}-\frac{1}{2}\xi_xg^{xt}g_{tt,x}=\kappa \xi_x.
\end{equation}
Substitution of the relevant values for the components of the metric tensor corresponding to the Gullstrand-Painlevé line element of Eq.~\eqref{Equation: Gullstrand-Painleve metric line element} into the above equation and evaluating the resultant expression at the event horizon gives the following surface gravity:
\begin{equation}
    \kappa = \pm \frac{du(x_h)}{dx}-\frac{dv(x_h)}{dx}.
\end{equation}

\subsection{Dirac equation in Rindler space-time}\label{Appendix: Dirac Equation in Rindler Space-Time}
Henceforth, we take $v$ to be a slowly varying position-dependent function $v(x)$ and restrict ourselves to the case where $u=1$, such that the Gullstrand-Painlevé line element of Eq.~\eqref{Equation: Gullstrand-Painleve metric line element} simplifies to
\begin{equation}
    ds^2=\left(1-v(x)^2\right)dt^2-2v(x)dtdx-dx^2,
    \label{Equation: Gullstrand-Painleve metric line element for u=1}
\end{equation}
where $v(x)$ satisfies $v(x_h)=\pm1$, and $v(x)\rightarrow0$ as $x\rightarrow \infty$ such that the metric of Eq.~\eqref{Equation: Gullstrand-Painleve metric line element for u=1} is \textit{asymptotically flat}. The coordinate transformation  $(t,x)\mapsto(\tilde{t},x)$ between Gullstrand-Painlev\'e coordinates $x^\mu=(t,x)$ and Schwarzschild coordinates $\tilde{x}^\mu=(\tilde{t},x)$, where
\begin{equation}
    t(\tilde{t},x)=\tilde{t}+\int dx\frac{v(x)}{1-v(x)^2},
\end{equation}
can be used to map the Gullstrand-Painlev\'e metric line element of Eq.~\eqref{Equation: Gullstrand-Painleve metric line element for u=1} to 
\begin{equation}
    ds^2=\left(1-v(x)^2\right){d\tilde{t}}^2-\frac{dx^2}{1-v(x)^2},
    \label{Equation: Schwarzschild metric line element}
\end{equation}
which is equivalent to the (1+1)-dimensional Schwarzschild metric when $v(x)=\sqrt{x_h/x}$. The metric of Eq~\eqref{Equation: Schwarzschild metric line element} contains a coordinate singularity at $v(x_h)=\pm1$ and is therefore only valid in the region $x>x_h$ that describes the black hole's exterior.

The function $f(x)=1-v(x)^2$ can be Taylor expanded about $x_h$ as
\begin{equation}
     f(x)\simeq 2\kappa(x-x_h)+O(x^2),
\end{equation}
where the metric's surface gravity $\kappa=-v'(x_h)$ has been identified. Hence, near the event horizon of the black hole, Eq.~\eqref{Equation: Schwarzschild metric line element} for the Schwarzschild metric line element can be expressed as
\begin{equation}
    ds^2=2\kappa(x-x_h){d\tilde{t}}^2-\frac{dx^2}{2\kappa(x-x_h)}.
    \label{Equation: Schwarzschild metric line element near horizon}
\end{equation}
A further coordinate transformation $(\tilde{t}, x)\mapsto(\tilde{t},\rho)$, where
\begin{equation}
    \rho^2=\frac{2(x-x_h)}{\kappa},
\end{equation}
can be employed to map Eq.~\eqref{Equation: Schwarzschild metric line element near horizon} for the line element of the Schwarzschild metric near the horizon to
\begin{equation}
    ds^2=(\kappa\rho)^2 d{\tilde{t}}^2-d\rho^2,
    \label{Equation: Rindler metric line element}
\end{equation}
which is the Rindler metric line element \cite{Rindler_1966, Rindler_1977} that describes an observer that is uniformly accelerating in Minkowski space-time with a constant proper acceleration $\kappa$. The coordinates $x^\mu_R=(\tilde{t}, \rho)$ will be referred to as Rindler coordinates. The Rindler metric possesses a coordinate singularity at $\rho=0$, which corresponds to the coordinate singularity at $x_h$ of the Schwarzschild metric. For this reason, we leave the Rindler metric undefined for the region $\rho<0$ \cite{Mertens_et_al_2022}. The zweibein corresponding to the Rindler metric line element of Eq.~\eqref{Equation: Rindler metric line element} is given by $e^a_\mu=\text{diag}(\kappa \rho, \ 1)$.

We now proceed with solving the Dirac equation for a spinor $\psi(\tilde{t},\rho)$ in a space-time described by the Rindler metric. The Dirac equation for a massive spinor $\psi$ on a general Riemann-Cartan space-time reads
\begin{gather}
    (i\gamma^\mu D_\mu-m)\psi=0,\\
    \frac{i}{2}\gamma^\mu\partial_\mu\psi+\frac{i}{2}\frac{1}{|e|}\partial_\mu(|e|\gamma^\mu\psi)-\frac{i}{2}\{\gamma^\mu, \omega_\mu \}\psi-m\psi=0,
    \label{Equation: Dirac Equation in General RC space-time}
\end{gather}
which in (1+1)-dimensions simplifies to
\begin{equation}
    \frac{i}{2}\gamma^ae^\mu_a\partial_\mu \psi+\frac{i}{2}\frac{1}{|e|}\gamma^a\partial_\mu(|e|e^\mu_a\psi)-m\psi=0
    \label{Equation: Dirac Equation in (1+1)D RC space-time}
\end{equation}
due to the fact that $\{\gamma^\mu,\omega_\mu\}\propto \{\gamma^a,[\gamma^b,\gamma^c]\}=0$. As the Rindler metric is stationary with respect to the Rindler time coordinate $\tilde{t}$, it admits a Killing vector $\zeta^\mu=\delta^\mu_{\tilde{t}}$ (or equivalently $\partial_{\tilde{t}}$) that is time-like in the region $\rho>0$. Hence, the energy $\epsilon=p_{\tilde{t}}$ of the spinor is a well-defined quantity that is conserved along the Rindler metric's geodesics \cite{Parker_Toms_2009, Huhtala_Volovik_2002}. Therefore, the Dirac equation can be solved via the separation of variables \cite{Huhtala_Volovik_2002, Fulling_1973, Fulling_Thesis}
\begin{equation}
    \psi(\tilde{t}, \rho)=e^{-i\epsilon_k\tilde{t}}\psi_{\epsilon_k}(\rho).
\end{equation}
Substitution of this solution, along with the components of the Rindler metric zweibein, into the Dirac equation of Eq.~\eqref{Equation: Dirac Equation in (1+1)D RC space-time} yields the following eigenvalue equation
\begin{equation}
    h_R\psi_{\epsilon_k}(\rho)= \epsilon_k\psi_{\epsilon_k}(\rho),
    \label{Equation: Rindler Eigenvalue equation}
\end{equation}
where 
\begin{equation}
h_R = i\kappa\rho\gamma^0\gamma^1 \partial_{\rho} +\frac{i}{2}\kappa\gamma^0\gamma^1-m\kappa\rho\gamma^0
\end{equation}
is a linear differential operator that is Hermitian under the conserved inner product 
\begin{equation}
    \langle\phi(\rho),\psi(\rho)\rangle=\int d\mu(\rho)\phi(\rho)^\dagger\psi(\rho),
    \label{Equation: Inner Product}
\end{equation}
and $d\mu(\rho)=d\rho$ is a measure of integration. That is, $h_R$ is Hermitian as it satisfies $\langle \phi,h_R\psi \rangle=\langle h_R\phi,\psi\rangle$ \cite{Fulling_1989}. Restricting ourselves to the case of a massless spinor and choosing the chiral representation of the $\gamma$ matrices, $\gamma^0=i\sigma^x$ and $\gamma^1=\sigma^y$, the Hermitian operator $h_R$ can be expressed as
\begin{equation}
    h_R=-i\kappa\rho\sigma^z\partial_\rho-\frac{i}{2}\kappa\sigma^z.
\end{equation}
This allows the eigenvalue equation given in Eq.~\eqref{Equation: Rindler Eigenvalue equation} to be expressed as the two following decoupled differential equations:
\begin{equation}
    \epsilon_k\psi^\pm_{\epsilon_k}(\rho)\pm i\kappa \rho\partial_\rho\psi^\pm_{\epsilon_k}(\rho)\pm\frac{i\kappa}{2}\psi^\pm_{\epsilon_k}(\rho)=0,
\end{equation}
where $\psi^{\pm}_{\epsilon_k}(\rho)$ denote the components of the spinor $\psi_{\epsilon_k}(\rho)$ that have solutions $\psi^\pm_{\epsilon_k}(\rho)=\mathcal{N}\rho^{\pm\frac{i\epsilon_k}{\kappa}-\frac{1}{2}}$ \cite{Mertens_et_al_2022, Denger_Horner_Pachos_2023}. Here, $\mathcal{N}$ is a normalisation constant that is chosen such that $\langle \psi_{\epsilon_k}, \psi_{\epsilon_q} \rangle=\delta(k-q)$, which, using the Dirac delta identity $\int^\infty_0x^{i\alpha-1}dx=2\pi\delta(\alpha)$, can be determined to be $\mathcal{N}=1/\sqrt{2\pi |\kappa|}$ \cite{Mertens_et_al_2022}. Thus, as a solution $\psi(\tilde{t},\rho)$ to the Dirac equation on the Rindler metric, we have $\psi(\tilde{t},\rho)=e^{-i\epsilon_k\tilde{t}}\psi_{\epsilon_k}(\rho)$, where
\begin{equation}
    \psi_{\epsilon_k}(\rho)=\frac{1}{\sqrt{2\pi|\kappa|}}\begin{pmatrix}
        \rho^{\frac{i\epsilon_k}{\kappa}-\frac{1}{2}}\\
        \rho^{-\frac{i\epsilon_k}{\kappa}-\frac{1}{2}}
    \end{pmatrix}.
\end{equation}

This solution, however, is not unique. For instance, $\psi_{\epsilon_k}(\rho)$ can be expressed as the linear combination of two orthonormal spinors $\varpsi^\pm_{\epsilon_k}(\rho)=\mathcal{N}\rho^{\pm\frac{i\epsilon_k}{\kappa}-\frac{1}{2}}u_\pm$, where $u_\pm$ are the eigenvectors of $\sigma^z$ with eigenvalues $\pm1$, respectively, which are also solutions to the Dirac equation and satisfy the eigenvalue equation $h_R\varpsi^\pm_{\epsilon_k}(\rho)=\epsilon_k\varpsi^\pm_{\epsilon_k}(\rho)$. Additionally, as we are considering a complex spinor field, there exists a further solution $\phi(\tilde{t},\rho)=C\psi^*(\tilde{t},\rho)=e^{i\epsilon_k\tilde{t}}\phi_{\epsilon_k}(\rho)$ to the Dirac equation \cite{Parker_Toms_2009}, which satisfies the eigenvalue equation $h_R\phi_{\epsilon_k}(\rho)=-\epsilon_k\phi_{\epsilon_k}(\rho)$ and the orthonormality condition $\langle \phi_{\epsilon_k}, \psi_{\epsilon_k}\rangle=0$. Here, $C$ is a charge conjugation operator, defined as $C^\dagger h_R C=-h_R^*$, and $\phi(\tilde{t},\rho)$ is the charge conjugate spinor of $\psi(\tilde{t},\rho)$ \cite{Parker_Toms_2009}. In the chiral representation, $C=\mathbb{I}$, hence $\phi_{\epsilon_k}(\rho)=\psi_{\epsilon_k}^*(\rho)$. Analogous to $\psi_{\epsilon_k}(\rho)$, $\phi_{\epsilon_k}(\rho)$ can be expressed via a linear combination of two orthonormal spinors $\varphi^\pm_{\epsilon_k}(\rho)=\mathcal{N}\rho^{\mp\frac{i\epsilon_k}{\kappa}-\frac{1}{2}}u_\pm$ that solve the Dirac equation and satisfy the eigenvalue equation $h_R\varphi^\pm_{\epsilon_k}(\rho)=-\epsilon_k \varphi^\pm_{\epsilon_k}(\rho)$. Thus, a general solution to the Dirac equation must be constructed with both $\psi_{\epsilon_k}(\rho)$ and $\phi_{\epsilon_k}(\rho)$ via an eigenfunction decomposition.

Assuming that the orthonormal basis formed by the eigenfunctions $\psi_{\epsilon_k}(\rho)$ of the Hermitian operator $h_R$ is complete, any function $F(\rho)$ within the Hilbert space $\mathcal{H}_R$ defined by the inner product of Eq.~\eqref{Equation: Inner Product} can be decomposed into a continuous sum of the eigenfunctions $\psi_{\epsilon_k}(\rho)$ as \cite{Fulling_1973, Fulling_Thesis, Fulling_1989}
\begin{equation}
    F(\rho)=\int d\mu(k)\tilde{f}(k)\psi_{\epsilon_k}(\rho),
    \label{Equation: Eigenfunction decomposition}
\end{equation}
where $d\mu(k)=dk$ is an integration measure of the continuous index $k$ that defines the inner product
\begin{equation}
    \langle \tilde{f}(k),\tilde{g}(k)\rangle=\int d\mu(k) \tilde{f}^\dagger(k)\tilde{g}(k)
\end{equation}
of a corresponding Hilbert space $\mathcal{L}^2(k)$ of functions $\tilde{f}(k)$, which is related to $\mathcal{H}_R$ via the eigenfunction decomposition of Eq.~\eqref{Equation: Eigenfunction decomposition} and its inverse \cite{Fulling_1973, Fulling_Thesis, Fulling_1989}
\begin{equation}
    \tilde{f}(k)=\int d\mu(\rho) \psi^\dagger_{\epsilon_k}(\rho)F(\rho).
    \label{Equation: Inverse eigenfunction decomposition}
\end{equation}
For a consistent correspondence between the two Hilbert spaces via the eigenfunction decomposition and its inverse, we require the normalisation conditions $\langle \psi_{\epsilon_k}(\rho), \psi_{\epsilon_q}(\rho)\rangle=\delta(k-q)$, where $\int d\mu(q)\delta(k-q)\tilde{f}(q)=\tilde{f}(k)$, along with $\langle \psi_{\epsilon_k}(\rho), \psi_{\epsilon_k}(\rho')\rangle=\delta(\rho-\rho')$, where \\
\noindent $\int d\mu(\rho')\delta(\rho-\rho')F(\rho')=F(\rho)$. 

Using the eigenfunction decomposition defined in Eq.~\eqref{Equation: Eigenfunction decomposition}, the most general solution $\psi(\tilde{t},\rho)$ to the Dirac equation in Rindler space-time can be decomposed into a continuous sum of the Dirac equation's spinor $\psi_{\epsilon_k}(\rho)$ and charge conjugate spinor $\phi_{\epsilon_k}(\rho)$ eigenfunctions. That is, 
\begin{equation}
    \psi(\tilde{t},\rho)=\int^\infty_0d\mu(k)\left( \tilde{f}(k)e^{-i\epsilon_k\tilde{t}}\psi_{\epsilon_k}(\rho)+\tilde{g}(k)e^{i\epsilon_k\tilde{t}}\phi_{\epsilon_k}(\rho) \right).
    \label{Equation: Dirac equation in Rindler metric mode decomposition in terms of Ek}
\end{equation}

From the eigenvalue equation of Eq.~\eqref{Equation: Rindler Eigenvalue equation}, it is evident that $\psi_{\epsilon_k}(\rho)$ is a positive energy solution when $\epsilon_k>0$ and a negative energy solution when $\epsilon_k<0$, with the converse being true for $\phi_{\epsilon_k}(\rho)$. Thus, to ensure that the spinor $\psi_{\epsilon_k}(\rho)$ and its charge conjugate $\phi_{\epsilon_k}(\rho)$ always correspond to positive and negative energy solutions, respectively, we define
\begin{equation}
    k=\begin{cases}
    \epsilon_k, & \text{if }\epsilon_k\geq0\\
    -\epsilon_k, &\text{if }\epsilon_k<0
    \end{cases}, \ \text{and} \
    u_k=\begin{cases}
        u_+, & \text{if }k\geq0 \\
        u_-, & \text{if }k<0
    \end{cases},
\end{equation}
such that the general solution to the Dirac equation in Eq.~\eqref{Equation: Dirac equation in Rindler metric mode decomposition in terms of Ek} can be expressed as \cite{Denger_Horner_Pachos_2023, Mertens_et_al_2022}
\begin{equation}
    \psi(\tilde{t}, \rho)=\int^\infty_{-\infty}d\mu(k) \left( \tilde{f}(k)e^{-i|k|\tilde{t}}\psi_k(\rho) + \tilde{g}(k)e^{i|k|\tilde{t}}\phi_k(\rho) \right), 
    \label{Equation: Dirac equation in Rindler metric mode decomposition in terms of k}
\end{equation}
and the eigenfunctions $\psi_k(\rho)=\mathcal{N}\rho^{\frac{ik}{\kappa}-\frac{1}{2}}u_k$ and $\phi_k(\rho)=\mathcal{N}\rho^{-\frac{ik}{\kappa}-\frac{1}{2}}u_k$ are always positive and negative energy solutions, respectively, satisfying the eigenvalue equations
\begin{equation}
\begin{split}
    h_R\psi_k(\rho)=\begin{cases}
        k\psi_k(\rho), & \text{for }k\geq 0\\
        -k\psi_k(\rho), & \text{for }k<0
    \end{cases}, \\
    h_R\phi_k(\rho)=\begin{cases}
        -k\phi_k(\rho), & \text{for }k\geq 0\\
        k\phi_k(\rho), & \text{for }k<0
    \end{cases}.
\end{split}
    \label{Equation: Eigenvalue equation of Dirac equation solutions in terms of k}
\end{equation}
As before, $\psi_k(\rho)$ and $\phi_k(\rho)$ satisfy the orthonormality conditions $\langle \psi_k(\rho), \phi_q(\rho)\rangle=0$ and $\langle \psi_k(\rho), \psi_q(\rho)\rangle=\langle \phi_k(\rho), \phi_q(\rho)\rangle=\delta(k-q)$.

\subsection{Quantum field theory of spinor in Rindler space-time}
To quantise the solution of Eq.~\eqref{Equation: Dirac equation in Rindler metric mode decomposition in terms of k} to the Dirac equation in Rindler space-time, $\psi(\tilde{t},\rho)$ and $\pi(\tilde{t},\rho)$ are promoted to Hermitian operators $\hat{\psi}(\tilde{t},\rho)$ and $\hat{\pi}(\tilde{t},\rho)$ and required to satisfy the equal-time canonical anti-commutation relations of Eq.~\eqref{Equation: Canonical Anti-Commutation Relations (Canonical Momentum)}, where the canonical momentum conjugate to $\hat{\psi}(\tilde{t},\rho)$ is given by $\hat{\pi}(\tilde{t},\rho)=-\frac{i}{2}\hat{\psi}^\dagger(\tilde{t},\rho)$ for the Rindler metric and by Eq.~\eqref{Equation: Canonical Momentum for Dirac Spinor} for a general Riemann-Cartan space-time. In quantising the Dirac field, the functions $\tilde{f}(k)$ and $\tilde{g}(k)$ are also promoted to operators $\hat{f}(k)$ and $\hat{g}(k)$. Using the inverse of the eigenfunction decomposition defined in Eq.~\eqref{Equation: Inverse eigenfunction decomposition}, it can be verified that these operators can be expressed in terms of the Cauchy initial data, $\hat{\psi}(0,\rho)$ and $\hat{\pi}(0,\rho)$, as $\hat{f}(k)=\langle \hat{\psi}_k(\rho), \hat{\psi}(0,\rho)\rangle$ and $\hat{g}(k)=\langle \hat{\phi}_k(\rho), \hat{\psi}(0,\rho)\rangle$ \cite{Parker_Toms_2009}. Using these expressions and the orthonormality conditions of $\psi_k(\rho)$ and $\phi_k(\rho)$, the following anti-commutation relations can be derived: \cite{Parker_Toms_2009}
\begin{align}
\begin{split}
     \{\hat{f}(k),\hat{f}^\dagger(q)\}=\{&\hat{g}(k),\hat{g}^\dagger(q)\}=\delta(k-q),\\
    \{\hat{f}(k),\hat{f}(q)\}&=\{\hat{g}(k),\hat{g}(q)\}=0,
\end{split}
\label{Equation: Anti-commutation relations of f(k) and g(f)}
\end{align}
which are those satisfied by typical creation and annihilation operators.

It can be explicitly shown that $\hat{f}(k)$ and $\hat{g}^\dagger(k)$ are creation and annihilation operators that act to raise and lower the energy of the Dirac field by discrete values. To do this, the Hamiltonian of the Dirac spinor on the Rindler space-time must be expressed in terms of $\hat{f}(k)$ and $\hat{g}^\dagger(k)$. From \eqref{Equation: Hamiltonian of Spinor on Riemann-Cartan Space-time} for the Hamiltonian of a spinor on a stationary Riemann-Cartan space-time, the Hamiltonian of the spinor on the Rindler space-time will be 
\begin{equation}
    \hat{H}=\int d\rho \ \hat{\pi}(\tilde{t},\rho)\partial_t\hat{\psi}(\tilde{t}, \rho),
\end{equation}
which, due to the Hamiltonian's Hermiticity, can be written as
\begin{equation}
    \hat{H}=-\frac{i}{2}\int d\rho (\partial_t\hat{\psi}^\dagger(\tilde{t},\rho))\hat{\psi}(\tilde{t},\rho).
\end{equation}
Substitution of Eq.~\eqref{Equation: Dirac equation in Rindler metric mode decomposition in terms of k} for the general solution to the Dirac equation in a Rindler space-time into the above Hamiltonian, and making use of the orthonormality conditions of $\psi_k(\rho)$ and $\phi_k(\rho)$, yields \cite{Parker_Toms_2009}
\begin{equation}
    \hat{H}=\frac{|k|}{2}\int d\mu(k) \left( \hat{f}^\dagger(k)\hat{f}(k)- \hat{g}^\dagger(k)\hat{g}(k)\right).
    \label{Equation: Hamiltonian of Dirac Spinor in Rindler Space-Time in terms of f(k) and g(k)}
\end{equation}
From this Hamiltonian and the anti-commutation relations of the operators $\hat{f}(k)$ and $\hat{g}(k)$ given in Eq.~\eqref{Equation: Anti-commutation relations of f(k) and g(f)}, the following commutation relations between $\hat{f}(k)$, $\hat{g}(k)$ and the Hamiltonian can be obtained:
\begin{gather}
    [\hat{H},\hat{f}^\dagger(k)]=\frac{|k|}{2}\hat{f}^\dagger(k), \ \ \ [\hat{H},\hat{f}(k)]=-\frac{|k|}{2}\hat{f}(k), \\
    [\hat{H},\hat{g}^\dagger(k)]=-\frac{|k|}{2}\hat{g}^\dagger(k), \ \ \ [\hat{H},\hat{g}(k)]=\frac{|k|}{2}\hat{g}(k),
\end{gather}
which, in conjunction with the eigenvalue equations of $\psi_k(k)$ and $\phi_k(k)$ given in Eq.~\eqref{Equation: Eigenvalue equation of Dirac equation solutions in terms of k}, can be used to deduce that $\hat{f}^\dagger(k)$ and $\hat{f}(k)$ are operators that act to raise and lower the energy of the quantum field by creating and annihilating quanta $\psi_k$ of energy $|k|/2$, respectively, whilst $\hat{g}^\dagger(k)$ and $\hat{g}(k)$ act to lower and raise the energy of the quantum field by creating and annihilating quanta $\phi_k$ of energy $-|k|/2$, respectively. To be in line with standard notation, we re-label $\hat{f}^\dagger(k), \hat{f}(k)\mapsto \hat{a}^\dagger_k,\hat{a}_k$, and $\hat{g}^\dagger(k), \hat{g}(k)\mapsto \hat{b}_k,\hat{b}^\dagger_k$. Therefore, the general solution to the Dirac equation in Rindler space-time is \cite{Denger_Horner_Pachos_2023, Mertens_et_al_2022}
\begin{align}
    \hat{\psi}(\tilde{t}, \rho)=\int^\infty_{-\infty}d\mu(k) \left( \hat{a}_ke^{-i|k|\tilde{t}}\psi_k(\rho) + \hat{b}^\dagger_ke^{i|k|\tilde{t}}\phi_k(\rho) \right),
    \label{Equation: Mode Decomposition Dirac Equation in Rindler Space-Time}
\end{align}
where $\hat{a}_k=\langle\psi_k(\rho),\psi(0,\rho)\rangle$, $\hat{b}^\dagger_k=\langle\phi_k(\rho),\psi(0,\rho)\rangle$, and $\{\hat{a}_k,\hat{a}^\dagger_q \}=\{\hat{b}_k,\hat{b}^\dagger_q \}=\delta(k-q)$, whilst all other anti-commutation relations vanish.

\subsection{Quantum field theory of spinor in Minkowski space-time}
To derive the Fulling-Davies-Unruh effect for a Rindler observer near the event horizon of the black hole, we are concerned with determining the vacuum expectation value of the Minkowski vacuum as seen by the Rindler observer \cite{Carroll_2019}. Therefore, a general solution to the Minkowski space-time Dirac equation must also be obtained, and a quantum field theory for a spinor in a Minkowski space-time constructed. In $(1+1)$-dimensions, the Minkowski metric line element is 
\begin{equation}
    ds^2=dT^2-dX^2,
    \label{Equation: Minkowski Metric Line Element}
\end{equation}
where the Minkowski coordinates $X^\mu=(T,X)$ are related to the Rindler coordinates $x_R^\mu=(\tilde{t},\rho)$ via 
\begin{equation}
    T=\rho \sinh(\kappa \tilde{t}), \ \ \ X=\rho \cosh(\kappa \tilde{t}).
\end{equation}
In Minkowski space-time, as the zweibein is trivially given by $e^a_\mu=\mathbb{I}$, the Dirac equation, as is given in Eq.~\eqref{Equation: Dirac Equation in General RC space-time}, simplifies to
\begin{equation}
    \left( i\gamma^ae^\mu_a\partial_\mu -m \right)\Psi(T,X)=0
    \label{Equation: Dirac Equation in Minkoski Spacetime}
\end{equation}
for the two-component spinor $\Psi(T,X)$. Analogous to the Rindler metric, as the Minkowski metric is stationary with respect to the Minkowski time coordinate $T$, the vector $K^\mu=\delta^\mu_T$ (or equivalently $\partial_T$) is a Killing vector that is time-like for all of the Minkowski space-time. Thus, the energy $E$ of the spinor is a well-defined quantity that is conserved along the Minkowski metric's geodesics \cite{Parker_Toms_2009, Huhtala_Volovik_2002}, and the Dirac equation can be solved via the separation of variables
\begin{equation}
    \Psi(T,X)=e^{-iE_KT}\Psi_{E_K}(X),
\end{equation}
where $\Psi_{E_K}(X)=(\Psi_{E_K}^+(X) \ \Psi_{E_K}^-(X))^T$ is a two-component spinor that depends only on $X$. Therefore, the Minkowski space-time Dirac equation can be expressed as the following eigenvalue equation:
\begin{equation}
    h_M\Psi_{E_K}(X)=E_K\Psi_{E_K}(X),
    \label{Equation: Minkowski Eigenvalue Equation}
\end{equation}
where $h_M=i\gamma^0\gamma^1\partial_X +m\gamma^0$ is a linear differential operator that is Hermitian under the conserved inner product 
\begin{equation}
    \langle \Phi(X),\Psi(X)\rangle=\int d\mu(X)\Phi^\dagger (X)\Psi(X),
    \label{Equation: Inner Product Minkowski space-time}
\end{equation}
and $d\mu(X)=dX$ is a measure of integration. That is, $h_M$ satisfies $\langle \Phi,h_M\Psi\rangle=\langle h_M\Phi,\Psi\rangle$ \cite{Fulling_1989}. Once again restricting ourselves to a massless spinor and choosing the chiral representation of the $\gamma$ matrices, the Hermitian operator $h_M$ can be expressed as $h_M=-u\sigma^z\partial_X\Psi_{E_K}(X)$. Therefore, the eigenvalue equation corresponds to the two following decoupled differential equations:
\begin{equation}
    E_K\Psi_{E_K}^\pm(X)\pm i\partial_X\Psi_{E_K}^\pm(x)=0,
\end{equation}
where $\Psi_{E_K}^\pm(x)$ denote the components of the spinor $\Psi_{E_K}(x)$ that have the solutions 
$\Psi_{E_K}^\pm(X)=\mathcal{N}e^{\pm i E_K X}$, where $\mathcal{N}=1/\sqrt{2\pi}$ is a normalisation constant chosen such that $\langle \Psi_{E_K}(X),\Psi_{E_Q}(X)\rangle=\delta(K-Q)$. Hence, as a solution to the Dirac equation in Minkowski space-time, we have
\begin{equation}
    \Psi(T,X)=\frac{1}{\sqrt{2\pi}}e^{-iE_K T}\begin{pmatrix} e^{iE_KX} \\ e^{-iE_KX} \end{pmatrix}
\end{equation}
The charge conjugate solution $\Phi(T,X)=C\Psi^*(T,X)=e^{iET}\Phi_{E_K}(X)$ to the Minkowski space-time Dirac equation, which satisfies the eigenvalue equation $h_M\Phi_{E_K}(x)=-E_K\Phi_{E_K}(X)$ and the orthonormality condition $\langle \Phi_{E_K}, \Psi_{E_K}\rangle=0$, will be given by $\Phi(T,X)=\Psi^*(T,X)$.

Again, assuming that the orthonormal basis formed by the eigenfunctions $\Psi_{E_K}(X)$ of the Hamiltonian $h_M$ is complete, any function $F(X)$ within the Hilbert space $\mathcal{H}_M$ defined by the conserved inner product of Eq.~\eqref{Equation: Inner Product Minkowski space-time} can be decomposed into a continuous sum of eigenfunctions $\Psi_{E_K}(X)$ as \cite{Fulling_1973, Fulling_1989, Fulling_Thesis}
\begin{equation}
    F(X)=\int d\mu(K)\tilde{f}(K)\Psi_{E_K}(X),
    \label{Equation: Eigenfunction Decomposition in Minkowski Space-Time}
\end{equation}
where $d\mu(K)=dK$ is an integration measure of the continuous index $K$ that defines the conserved inner product 
\begin{equation}
    \langle \tilde{f}(K),\tilde{g}(K)\rangle=\int\mu(K)\tilde{f}^\dagger(K)\tilde{g}(K),
\end{equation}
of a corresponding Hilbert space $\mathcal{L}^2(K)$ of functions $\tilde{f}(K)$, which is related to $\mathcal{H}_M$ via the eigenfunction decomposition of Eq.~\eqref{Equation: Eigenfunction Decomposition in Minkowski Space-Time} and its inverse \cite{Fulling_1973, Fulling_1989, Fulling_Thesis}
\begin{equation}
    \tilde{f}(K)=\int d\mu(X)\Psi_{E_K}^\dagger(X)F(X).
    \label{Equation: Inverse Eigenfunction Decomposition in Minkowski Space-Time}
\end{equation}
To consistently relate the two Hilbert spaces via the eigenfunction decomposition and its inverse, we require the normalisation conditions $\langle \Psi_{E_K}(X), \Psi_{E_Q}(X)\rangle=\delta(K-Q)$, along with $\int d\mu(Q)\delta(K-Q)\tilde{f}(Q)=\tilde{f}(K)$, and $\langle \Psi_{E_K}(X), \Psi_{E_K}(X')\rangle=\delta(X-X')$, with the condition that $\int d\mu(X')\delta(X-X')F(X')=F(X)$ \cite{Fulling_1973, Fulling_1989, Fulling_Thesis}.

As in the Rindler case, from the eigenfunction decomposition of Eq.~\eqref{Equation: Eigenfunction Decomposition in Minkowski Space-Time}, the most general solution $\Psi(T,X)$ of the Minkowski space-time Dirac equation can be decomposed into a continuous sum of the Dirac equation's spinor $\Psi_{E_K}(X)$ and charge conjugate spinor $\Phi_{E_K}(X)$ eigenfunctions. That is, 
\begin{equation}
\begin{split}
    \Psi(T,X)\!=\!\int^\infty_{-\infty}d\mu(K)\Bigl(&\tilde{f}(K)e^{-i|K|T}\Psi_K(X) \\
    &\  \ +\!\tilde{g}(K)e^{i|K|T}\Phi_K(X)\Bigr),
\end{split}
\label{Equation: General Solution to Dirac Eq in Minkowski Space-Time in Terms of E}
\end{equation}
where, to once again ensure that the spinor $\Psi_{E_K}$ and its charge conjugate $\Phi_{E_K}$ always correspond to positive and negative energy solutions, respectively, we have defined
\begin{equation}
    K=\begin{cases}
        E_K,& \text{if }E_K\geq0 \\
        -E_K,& \text{if }E_K<0
    \end{cases}, \ \ \ 
    u_K=\begin{cases}
        u_+,& \text{if } K\geq0 \\
        u_-,& \text{if} K<0
    \end{cases}
\end{equation}
and the eigenfunctions $\Psi_K(X)=\mathcal{N}e^{-iKX}u_K$ and $\Phi_K(X)=\mathcal{N}e^{iKX}u_K$.

Once again, this solution to the Dirac equation for a spinor in Minkowski space-time can be quantised by promoting $\Psi(T,X)$ and $\Pi(T,X)=-\frac{i}{2}\Psi^\dagger(T,X)$ to Hermitian operators and imposing the equal-time canonical anti-commutations of Eq.~\eqref{Equation: Canonical Anti-Commutation Relations (Canonical Momentum)}. In quantising the Dirac field in Minkowski space-time, the functions $\tilde{f}(K)$ and $\tilde{g}(K)$ are promoted to the operators $\hat{f}(K)$ and $\hat{g}(K)$, respectively. Using the inverse eigenfunction decomposition defined in \eqref{Equation: Inverse Eigenfunction Decomposition in Minkowski Space-Time}, it can be verified that the operators $\hat{f}(K)$ and $\hat{g}(K)$ can be expressed in terms of the Cauchy initial data, $\hat{\Psi}(0,X)$ and $\hat{\Pi}(0,X)$, as $\hat{f}(K)=\langle \hat{\Psi}_K(X),\hat{\Psi}(0,X)\rangle$ and $\hat{g}(K)=\langle \hat{\Phi}_K(X),\hat{\Psi}(0,X)\rangle$. From these expressions and the orthonormality conditions of $\hat{\Psi}_K(X)$ and $\hat{\Phi}_K(X)$, the following anti-commutation relations can be derived \cite{Parker_Toms_2009}:
\begin{gather}
    \{ \hat{f}(K), \hat{f}^\dagger(Q) \}=\{ \hat{g}(K), \hat{g}^\dagger(Q) \}=\delta(K-Q), \\
    \{ \hat{f}(K), \hat{f}(Q) \}=\{ \hat{g}(K), \hat{g}^\dagger(Q) \}=0,
\end{gather}
which are the anti-commutation relations satisfied by typical creation and annihilation operators.

Analogous to the Rindler space-time case, it then follows that $\hat{f}^\dagger(K)$ and $\hat{g}(K)$ are operators that act to raise and lower the energy of the quantum field by creating and annihilating quanta $\Psi_K$ of energy $|K|/2$, respectively, whilst $\hat{g}^\dagger(K)$ and $\hat{g}(K)$ are operators that act to lower and raise the energy of the quantum field by creating and annihilating quanta $\Phi_K$ of energy $-|K|/2$, respectively. Thus, after re-labelling $\hat{f}(K), \hat{f}(K)\mapsto\hat{A}^\dagger(K),\hat{A}_K$, and $\hat{g}^\dagger(K), \hat{g}(K)\mapsto\hat{B}_K,\hat{B}^\dagger_K$, the general solution to the Dirac equation in Minkowski space-time is \cite{Mertens_et_al_2022}
\begin{equation}
\begin{split}
    \hat{\Psi}(T,X)\!=\!\int^\infty _{-\infty} d\mu(K)\Bigl( &\hat{A}_K e^{-i|K|T}\Psi_K(X) \\ & \ \ +\!\hat{B}^\dagger_Ke^{i|K|T}\Phi_K(X)\Bigr),
\end{split}
    \label{Equation: General Solution to Dirac Equation in Minkowski Space-Time}
\end{equation}
where $\hat{A}_K=\langle \Psi_K, \Psi(0,X)\rangle$, $\hat{B}^\dagger_K=\langle\Phi_K, \Psi(0,X)\rangle$, and $\{\hat{A}_K,\hat{A}^\dagger_Q\}=\{\hat{B}_K,\hat{B}^\dagger_Q\}=\delta(K-Q)$, whilst all other anti-commutation relations vanish.

\subsection{Bogoliubov transformation and Hawking effect}\label{Appendix: Bogoloubov transformation and Hawking effect}
The Fulling-Davies-Unruh effect originates in the fact that the vacuum states of a Minkowski observer and a Rindler observer are not equivalent \cite{Wald_1994, Unruh_1975}. An observer in a Minkowski space-time will define their vacuum state as the state $|0_M\rangle$ that satisfies $\hat{A}_K|0_M\rangle=\hat{B}_K|0_M\rangle=0$ $\forall K$, where $\hat{A}_K$ and $\hat{B}_K$ are the particle and anti-particle annihilation operators, respectively, associated with the general solution of Eq.~\eqref{Equation: General Solution to Dirac Equation in Minkowski Space-Time}. On the other hand, a Rindler observer will define their vacuum state as the state $|0_R\rangle$ that satisfies $\hat{a}_k|0_R\rangle=\hat{b}_k|0_R\rangle=0$ $\forall k$, where $\hat{a}_k$ and $\hat{b}_k$ are the particle and anti-particle annihilation operators, respectively, associated with the general solution of Eq.~\eqref{Equation: Mode Decomposition Dirac Equation in Rindler Space-Time}. Generally, the vacuum states $|0_M\rangle$ and $|0_R\rangle$ are not equivalent, as the quantum field theories for the Minkowski and Rindler observers have been constructed by quantising general solutions to the Dirac equation that were obtained for different coordinate systems. Nevertheless, the two quantum field theories should be equivalent when the Minkowski and Rindler coordinate systems are equivalent \cite{Fulling_Thesis, Fulling_1973}. This occurs at $\tilde{t}=T=0$, $\rho=X$; thus, at $\tilde{t}=T=0$, we have $\hat{\psi}(0,\rho)=\hat{\Psi}(0,X)$ and $\hat{\pi}(0,\rho)=\hat{\Pi}(0,X)$ \cite{Fulling_1973}. As the creation and annihilation operators are determined by each theory's Cauchy initial data $\hat{\psi}(0,\rho)$ and $\hat{\Psi}(0,X)$, their equivalence at $\tilde{t}=T=0$ allows each theory's creation and annihilation operators to be expressed in terms of the other's. That is,
\begin{equation}
\begin{split}
\hat{a}_k\!&=\!\langle \hat{\psi}_k, \hat{\psi}(0,\rho)\rangle \\
         &=\!\int d\mu(K) \! \left( \hat{A}_K \langle \hat{\psi}_k,\hat{\Psi}_K\rangle\!+\!\hat{B}_K^\dagger \langle \hat{\psi}_k,\hat{\Phi}_K\rangle\right)
\end{split}
\label{Equation: Bogoliubov Transformation of Annihilation Operator}
\end{equation}
and
\begin{equation}
\begin{split}
\hat{b}^\dagger_k\!&=\!\langle \hat{\phi}_k, \hat{\psi}(0,\rho) \rangle \\
                &=\!\int d\mu(K)\!\left( \hat{A}_K \langle \hat{\phi}_k,\hat{\Psi}_K\rangle\!+\!\hat{B}_K^\dagger \langle \hat{\phi}_k,\hat{\Phi}_K\rangle \right),
\end{split}
\label{Equation: Bogoliubov Transformation of Creation Operator}
\end{equation}
where the equivalence of the two integration measures $d\mu(\rho)$ and $d\mu(X)$ at $\tilde{t}=T=0$ is used to compute the inner product between the Minkowski and Rindler eigenstates. Eqs.~\eqref{Equation: Bogoliubov Transformation of Annihilation Operator} and \eqref{Equation: Bogoliubov Transformation of Creation Operator} each give a Bogoliubov transformation between the Minkowski and Rindler creation and annihilation operators \cite{Denger_Horner_Pachos_2023, Mertens_et_al_2022, Fulling_1973}. Using these Bogoliubov transformations and the fact that $\hat{A}_K|0_M\rangle=\langle 0_M|\hat{A}^\dagger_K=0$, the vacuum expectation value of the Minkowski vacuum, as seen by the Rindler observer, can be expressed as
\begin{equation}
    \langle0_M|\hat{a}^\dagger_k \hat{a}_q|0_M\rangle =\int d\mu(K) \langle\hat{\phi}_k,\hat{\Psi}_K\rangle\langle \hat{\psi}_q, \hat{\Phi}_K \rangle,
\end{equation}
where the inner products, $\langle\hat{\phi}_k,\hat{\Psi}_K\rangle$ and $\langle \hat{\psi}_q, \hat{\Phi}_K \rangle$, can be explicitly calculated as follows: \cite{Mertens_et_al_2022}
\begin{align}
    \langle \hat{\psi}_q, \hat{\Phi}_K \rangle&=\frac{u^T_qu_K}{2\pi\sqrt{|\kappa|}}\int^\infty_0 dX \left( X^{-\frac{iq}{\kappa}-\frac{1}{2}}e^{-iKX}\right), \\
    &=\frac{\delta_{qK}}{2\pi\sqrt{|\kappa|}}e^{-\frac{\pi q}{2\kappa}-\frac{i\pi}{4}}K^{\frac{iq}{\kappa}-\frac{1}{2}}\Gamma\left(\frac{1}{2}-\frac{iq}{\kappa}\right),
\end{align}
where denotes the $\Gamma(z)$ is the gamma function and, in the final equality, the identity $i=e^{\frac{i\pi}{2}}$ was used;
\begin{align}
    \langle \hat{\phi}_k, \hat{\Psi}_K \rangle&=\frac{u^T_ku_K}{2\pi\sqrt{|\kappa|}}\int^\infty_0dX\left(X^{\frac{ik}{\kappa}-\frac{1}{2}}e^{iXK} \right),  \\
    &=\frac{\delta_{kK}}{2\pi\sqrt{|\kappa|}}e^{-\frac{\pi k}{2\kappa}+\frac{i\pi}{4}}K^{-\frac{ik}{\kappa}+\frac{1}{2}}\Gamma^*\left(\frac{1}{2}-\frac{ik}{\kappa}\right),
\end{align}
where $\Gamma^*(z)=\Gamma(z^*)$ and, in the final equality, the identity $-i=e^{-\frac{i\pi}{2}}$ was used. Hence, the vacuum expectation value of the Minkowski vacuum, as seen by the Rindler observer, can be found to be
\begin{align}
\begin{split}
    \langle0_M|\hat{a}^\dagger_k\hat{a}_q|0_M\rangle&=\frac{\delta_{kq}}{2\pi}e^{-\frac{\pi}{2\kappa}(k+q)}\Gamma^*\left(\frac{1}{2}-\frac{ik}{\kappa}\right)\\
    & \qquad \qquad \Gamma\left(\frac{1}{2}-\frac{iq}{\kappa}\right)\delta(k-q), 
\end{split}\\
    \langle0_M|\hat{a}^\dagger_k\hat{a}_k|0_M\rangle&=\frac{\delta(0)}{2}e^{-\frac{\pi k}{\kappa}}\text{sech}\left(\frac{\pi k}{\kappa}\right), \\
    &=\frac{\delta(0)}{e^{2\pi k/\kappa}+1},
    \label{Equation: Minkowski Vacuum Expectation Value as Seen by Rindler Observer}
\end{align}
where, in the second equality, the identity $|\Gamma(ix+\frac{1}{2})|^2=\pi \text{sech}(\pi x)$ has been used. Eq.~\eqref{Equation: Minkowski Vacuum Expectation Value as Seen by Rindler Observer} for the Minkowski vacuum expectation value, as seen by the Rindler observer, is a Fermi-Dirac distribution for fermions in thermal equilibrium with a temperature $T_H=\kappa/2\pi$. That is, an observer accelerating with a constant proper acceleration $\kappa$ will experience their vacuum state to be populated by thermal radiation with a temperature $T_H$ \cite{Fulling_1973, Davies_1975, Unruh_1975}; this is the Fulling-Davies-Unruh effect. As the Schwarzschild metric can be approximated by the Rindler metric near the event horizon, an observer near the event horizon of a Schwarzschild black hole will experience the Fulling-Davies-Unruh effect \cite{Wald_1994, Wald_1984a}. 

\vspace{10pt}
\section{Fermion zero-mode entropy of (1+1)D black hole exterior}\label{Appendix: Entropy of Black Hole Exterior}
\setcounter{equation}{0}
In this appendix, we extend the analysis of Sec.~\ref{Section: Entropy of (1+1)D Black Hole Fermions} to derive the fermion zero-mode entropy of both the black hole's interior and exterior regions. This derivation is analogous to that of Sec.~\ref{Section: Entropy of (1+1)D Black Hole Fermions} and involves determining the density of states of Eq.~\eqref{Equation: Density of States for General Couplings} for both the interior region, which is bound by $[a_c,x_h-a_c]$, and a portion of the exterior space-time, which we take to be bound by $[x_h+a_c, s]$, where $s\geq x_h+a_c$. To prevent an infinite number of degrees of freedom, from points on either side of the event horizon with a separation of less than $a_c$, from contributing to the density of states, the lower limit of $s$ is restricted to a distance $a_c$ from the event horizon \cite{Bombelli_et_al_1986, Callan_1994, Holzhey_1994}. 

Considering the emergence of the two zero-modes at momenta $p_\pm$ in the interior region, along with the two zero-modes at $p_0$ that exist throughout the entire space-time, the density of states of Eq.~\eqref{Equation: Density of States Integral} for the interior and exterior regions will be given by
\begin{widetext}
\begin{equation}
\begin{split}
    N(0)=\frac{N_F}{2\pi \hbar}\int^{x_h-a_c}_{a_c}dx \Biggl(\frac{1}{|1+v(x)|} &+ \frac{1}{|1-v(x)|}+\left|\frac{1}{\left(1-v(x)\right)}-\frac{1}{\left(1+v(x)\right)} \right| \Biggr) \\
    &\qquad \qquad  +\frac{\tilde{N}_F}{2\pi \hbar}\int^{s}_{x_h+a_c}dx \left(\frac{1}{|1+v(x)|}+ \frac{1}{|1-v(x)|}  \right),
\end{split}
\end{equation}
where $\tilde{N}_F$ denotes the integer number of massless fermionic fields in the exterior region. Taking the coupling corresponding to the space-time of a Schwarzschild black hole, $v(x)=\sqrt{x_h/x}$, and working in the limit $0< a_c\ll x_h$, this density of states can be solved to yield
\begin{equation}
    N(0)=\frac{2N_F x_h}{\pi \hbar}\left( \ln\left(\frac{4x_h}{a_c}\right) -2\right)+\frac{\tilde{N}_Fx_h}{\pi \hbar}\left(\frac{s}{x_h}+\ln\left| \frac{s-x_h}{a_c}\right| -1\right),
\end{equation}
\end{widetext}
where the term preceded by $N_F$ corresponds to the density of states of the entire black hole interior, and that preceded by $\tilde{N}_F$ the density of states of the exterior space-time. 

Assuming that the fermions in the black hole's interior and exterior regions are well described by a thermal Gibbs state, their thermal energy and entropy can be defined by Eqs.~\eqref{Equation: Thermal Energy of Fermions} and \eqref{Equation: Thermal Entropy of Fermions}, respectively. Once again, further assuming that the temperature associated with this thermal state is the Hawking temperature of Eq.~\eqref{Equation: Hawking Temperature of Chiral Spin-Chain Model}, the fermion zero-mode entropy of the black hole's interior and exterior will be given by
\begin{equation}
    \mathcal{S}(T_H)=\frac{N_F}{6} \ln\left(\frac{4x_h}{a_c}\right)+\frac{\tilde{N}_F}{12}\left( \frac{s}{x_h} +\ln\left| \frac{s-x_h}{a_c}\right|\right)+\mathcal{S}'_0
    \label{Equation: Entropy of (1+1)D Black Hole Fermion Zero-Modes Exterior (Appendix)}
\end{equation}
where $\mathcal{S}'_0=\mathcal{S}_0-\tilde{N}_F/12$. For the case where $s=x_h+a_c$, and in the limit $a_c\ll x_h$, Eq.~\eqref{Equation: Entropy of (1+1)D Black Hole Fermion Zero-Modes Exterior (Appendix)} for the fermion zero-mode entropy of the black hole's interior and exterior reduces to Eq.~\eqref{Equation: Entropy of (1+1)D Black Hole Fermion Zero-Modes at Horizon}, as expected.

\section{Conformal Invariance of Lattice Simulator}\label{Appendix: Conformal Invariance of Lattice Simulator}
In this appendix, we study the correspondence between a transformation of the mean-field Hamiltonian and a change in coordinates of the continuum limit Hamiltonian. This correspondence arises due to an equivalence between the choice of spinor renormalisation and a conformal scaling of the emergent metric. In the continuum limit, this transformation preserves the curvature of the underlying space-time provided that the couplings are also transformed accordingly; thus, scalar invariant quantities, such as the Ricci scalar, remain unchanged.

In deriving the continuum limit of the mean-field Hamiltonian of Eq.~\eqref{Equation: Mean-Field Hamiltonian}, the choice of zweibein defined by Eq.~\eqref{Equation: Zweibein of GP Metric} is not unique. One could alternatively define another set of coefficients $\tilde{e}^\mu_a$ that are related to $e^\mu_a$ via the transformation $e^\mu_a=C(u,v)\tilde{e}^\mu_a$, where, as we will see, $C(u,v)$ is a conformal scale factor. Following this transformation, the mean-field continuum limit Hamiltonian of Eq.~\eqref{Equation: MFT Hamiltonian in Momentum Space (Continuum)} can be expressed as 
\begin{equation}
    H=\int dp~C(u,v) \chi^\dagger(p)\tilde{e}^x_a \alpha^a p\chi(p),
\end{equation}
which, following an inverse Fourier transformation and, subsequently, a Legendre transformation, can be used to obtain the action
\begin{equation}
    S=i\int d^{1+1}x|\tilde{e}|\bar{\varphi}(x)\tilde{e}^\mu_a\gamma^a\overset{\leftrightarrow}{\partial}_\mu\varphi(x).
    \label{Equation: Conformally Scaled MFT Action}
\end{equation}
Here, the curved space-time spinor $\varphi$, which is related to the lattice spinor $\chi$ via the renormalisation $\chi=\sqrt{|\tilde{e}|/C}\varphi$ and satisfies the anti-commutation relations of Eq.~\eqref{Equation: Canonical Anti-Commutation Relations (Spinors)}, and its conjugate $\bar{\varphi}=\varphi^\dagger(x)\gamma^0$, have been defined.

The action of Eq.~\eqref{Equation: Conformally Scaled MFT Action} is, once again, that of a massless Dirac spinor $\varphi$ propagating freely on a (1+1)-dimensional Riemann-Cartan space-time, with the zweibein $\tilde{e}^\mu_a$ corresponding to the space-time metric line element
\begin{equation}
    d\tilde{s}^2=C(u,v)^2\left[\left(1-\frac{v^2}{u^2}\right)dt^2-\frac{2v}{u^2}dtdx -\frac{dx^2}{u^2} \right].
    \label{Equation: Line Element of Conformally Transformed GP metric}
\end{equation}
From the line element of Eq.~\eqref{Equation: Line Element of Conformally Transformed GP metric}, we see that $C(u,v)$ is a conformal scale factor that relates the Gullstrand-Painlevé line element of Eq.~\eqref{Equation: Gullstrand-Painleve metric line element} to that of Eq.~\eqref{Equation: Line Element of Conformally Transformed GP metric} via the conformal transformation $d\tilde{s}^2=C(u,v)^2ds^2$.

A unique case of the general curved space-time of Eq.~\eqref{Equation: Line Element of Conformally Transformed GP metric} can be obtained by taking $C(u)=\sqrt{u(x)}$ and working in the limit where $v\rightarrow 0$. In this case, the line element of Eq.~\eqref{Equation: Line Element of Conformally Transformed GP metric} reduces to 
\begin{equation}
    d\tilde{s}^2=u(x)dt^2-\frac{dx^2}{u(x)},
    \label{Equation: Schwarzschild Metric Line Element}
\end{equation}
which is that of the (1+1)-dimensional Schwarzschild metric. In this case, the corresponding lattice Hamiltonian is that of Eq.~\eqref{Equation: Mean-Field Hamiltonian} with a position-dependent nearest-neighbour coupling $u(x)$ and vanishing chirality term.

A further interesting case arises in the limit where $v\rightarrow0$ when taking $C=1$. Here, the line element of Eq.~\eqref{Equation: Line Element of Conformally Transformed GP metric} reduces to that of the (1+1)-dimensional Schwarzschild metric expressed in Lema{\^i}tre coordinates. If, then, one takes the nearest-neighbour coupling to the position and time-dependent coupling $u(t,x)=\sqrt{(1-kx^2)}/a(t)$, we obtain the line element
\begin{equation}
    d\tilde{s}^2=dt^2-a(t)^2\frac{dx^2}{1-kx^2},
\end{equation}
which is that of the (1+1)-dimensional Friedmann-Lemaître-Robinson-Walker (FLRW) metric \cite{Friedman_1922, Friedmann_1924, Lemaitre_1927, Lemaitre_1931, Robertson_1935, Robertson_1936_a, Robertson_1936_b, Walker_1937}, where $a(t)$ is the cosmological scale factor and $k$ is a constant that parametrises the metric's curvature.

\end{document}